\documentclass[singlecolumn,citeautoscript,superscriptaddress]{revtex4-1}  
\usepackage{amsmath,amssymb,mathrsfs,bm,feynmf,setspace,xspace,soul,empheq}  
\usepackage{graphicx,tikz,xparse}         
\usepackage[tight]{subfigure}      
\usepackage{color}  
 \usepackage{geometry}           
\usepackage[colorlinks=true]{hyperref}      
\hypersetup{
    bookmarks=true,         
    unicode=false,          
    pdftoolbar=true,        
    pdfmenubar=true,        
    pdffitwindow=false,     
    pdfstartview={FitH},    
    pdftitle={My title},    
    pdfauthor={Author},     
    pdfsubject={Subject},   
    pdfcreator={Creator},   
    pdfproducer={Producer}, 
    pdfkeywords={keyword1} {key2} {key3}, 
    pdfnewwindow=true,      
    colorlinks=true,       
    linkcolor=magenta, 
    citecolor=blue,        
    filecolor=magenta,      
    urlcolor=cyan           
} 

\newcommand{\ket}[1]{\left| #1\right\rangle}

\newcommand{\beq}{\begin{equation}}
\newcommand{\eeq}{\end{equation}}
\newcommand{\ba}{\begin{array}{ccc}}
\newcommand{\ea}{\end{array}}

\def\bea{\begin{eqnarray}}
\def\eea{\end{eqnarray}}
\newcommand{\bml}{\begin{multline}}

\newcommand{\eeqm}{\end{multline}}
\newcommand{\bsp}{\begin{split}}
\newcommand{\esp}{\end{split}}
\newcommand{\down}{\downarrow}
\newcommand{\up}{\uparrow}

\newcommand{\mc}{\mathcal}

\begin{document}  

\title{Gapless quantum spin chains: \\
multiple dynamics and conformal wavefunctions} 
\author{Xiao Chen}
\email{xchen@kitp.ucsb.edu} 
\affiliation{Kavli Institute for Theoretical Physics,
University of California at Santa Barbara, CA 93106, USA}

\author{Eduardo Fradkin}   
\email{efradkin@illinois.edu} 
\affiliation{Department of Physics and Institute for Condensed Matter Theory, University of Illinois at Urbana-Champaign, 1110 West Green Street,
Urbana, Illinois 61801-3080, USA} 

\author{William Witczak-Krempa}  
\email{w.witczak-krempa@umontreal.ca}   
\affiliation{D\'epartement de physique, Universit\'e de Montr\'eal, Montr\'eal (Qu\'ebec), H3C 3J7, Canada}

 \date{\today}  

\begin{abstract} 

We study gapless quantum spin chains with spin 1/2 and 1: the Fredkin and Motzkin models. Their entangled groundstates are known exactly but not 
their excitation spectra. We first express the groundstates in the continuum which allows for the
calculation of spin and entanglement properties in a unified fashion. Doing so, we uncover an emergent conformal-type symmetry, thus consolidating the connection 
to a widely studied family of Lifshitz quantum critical points in 2d. 
We then obtain the low lying excited states via large-scale DMRG simulations and find that 
the dynamical exponent is $z=3.2$ in both cases.
Other excited states show a different $z$, indicating that these models have multiple dynamics.  Moreover, we 
modify the spin-1/2 model by adding a ferromagnetic Heisenberg term, which changes the entire spectrum. 
We track the resulting non-trivial evolution of the dynamical exponents using DMRG.  
Finally, we exploit an exact map from the quantum Hamiltonian to the non-equilibrium dynamics of a classical spin chain to shed light on the quantum dynamics.   
\end{abstract} 
 
\maketitle    
 \date{\today}
\maketitle   
 
\tableofcontents  
\section{Introduction}   
Quantum critical systems display striking emergent phenomena such as universality in their static and dynamic properties.\cite{ssbook}
Realistic models are however often very challenging to study due to the presence of strong interactions.
This is why one-dimensional (1d) systems offer an ideal playground  
because they are often more tractable both numerically and analytically.   
For instance, many 1d quantum critical systems are described by conformal field theories (CFTs) at low energy. 
Such theories are highly constrained by symmetry alone.\cite{bigyellowbook,Fradkin-book}  
Even non-equilibrium dynamics following a quench can be studied in detail in 1d CFTs.\cite{CC06}  
A property that is useful to asses gapless quantum systems, both in and out of equilibrium, has come to the fore in recent years: the structure of entanglement.
For instance the bi-partite entanglement entropy (EE) quantifies the amount of quantum entanglement between two subsystems.
For 1d CFTs, the von Neumann EE has the form $S_{\rm vN}=\frac{c}{3}\log\frac{L_A}{\epsilon}$, where $L_A$ is the length of region $A$ (an interval embedded 
in the real line),
and $c$ is the central charge of the CFT.\cite{callan1994, Holzhey-1994, calabrese_entanglement_2004} 
$\epsilon$ is a short-distance cutoff. 
In addition, a related but cutoff-independent quantity, the mutual information between two well-separated intervals has a power law dependence on the separation.
The exponent depends on the scaling dimensions of local observables.\cite{Calabrese_MI_2009} 
These results demonstrate that the groundstates of CFTs are highly entangled. 

However, many gapless systems cannot be described by relativistic CFTs at low energy. 
An example that we shall investigate in this work
is the $S=1$ spin chain introduced by Bravyi et al: it describes spin exchange with bilinear and biquadratic interactions.\cite{Sergey2012}  
It is called the Motzkin model because the groundstate is the equal weight superposition of so-called Motzkin paths (more on this below).  
Although the groundstate has a large EE, the dynamical exponent that describes the low energy gapless excitations 
obeys the bound $z\geq 2$,\cite{Sergey2012, Movassagh_gap_2016} excluding the possibility that this model is described by a relativistic CFT. 
In addition, a recent large-scale DMRG study has shown that this model has different dynamical exponents for different excitations, 
thus exhibiting multiple dynamics at low energy.\cite{short-prep} 

A similar spin model with $S=1/2$ was proposed in Ref.~\onlinecite{DellAnna2016}. This model involves three-spin interaction 
and is related with Fredkin gates in the field of quantum computation, which explains why it is called the Fredkin model. 
Its groundstate EE also has logarithmic dependence for an interval embedded in a long chain. 
In this paper, we perform large-scale DMRG calculations to determine the excited states and show that this model also has $z>1$, meaning that it cannot be described by CFT, and that it displays multiple dynamics just as the Motzkin spin chain.  

Another reason why these spin models are interesting is because their Hamiltonians can be written as a sum of local projectors. 
The groundstate is annihilated by all the projectors, and has an energy exactly equal to zero. As such, these spin chains share a similar structure to 
the quantum dimer model (QDM) defined in 2d, and have potential applications in quantum computation due to their ``frustration-free'' nature.  
The QDM was introduced by Rokhsar and Kivelson to describe the low-lying singlet excitations in quantum antiferromagnets.\cite{Rokhsar1988} 
On the square lattice, the QDM model has a special parameter in its phase diagram, known as the Rokhsar-Kivelson (RK) point, 
where the critical groundstate is an equal weight superposition of all dimer configurations. This state admits an integer valued height representation 
defined mod 4, \cite{Baxterbook, Nienhuis1987} and in the continuum limit, the coarse-grained height field can be described by a  
free compact boson with a $z=2$ dispersion, 
and therefore the quantum wavefunction is conformally invariant in 2D space.\cite{Ardonne-2004} 
The RK form of the QDM Hamiltonian ensures that one can map the quantum dynamics to a non-equilibrium classical system governed by a Markovian master equation. Under this mapping, the groundstate coincides with the classical equilibrium distribution and the low energy excited state corresponds to the classical 
relaxation modes. Relying on this quantum-classical connection, Henley used classical Monte Carlo simulations to numerically study the dynamics in 
QDM and found that the dynamical exponent is $z=2$ \cite{henley_relaxation_1997,Moessner-2001}. This result suggests that the QDM at the RK point can be effectively described by the $z=2$ quantum Lifshitz model \cite{Ardonne-2004, Fradkin-book}. 

It should be noted, however, that it is not true that all lattice models which can be expressed as a sum of local projection operator must necessarily have $z=2$ dynamics. A case to the point is the 2d quantum eight vertex model of Ref. \cite{Ardonne-2004} which, as its classical analog, has two lines of fixed points where the 2D quantum system is critical. While along one of these critical lines, the six vertex line, the dynamics  indeed has $z=2$ dynamics, along the so-called Ashkin-Teller line using classical Monte Carlo simulations, Ref. \cite{Isakov2011} found a continuously varying  dynamical critical exponent. Here we will find a similar behavior in the Fredkin and Motzkin chains.

Keeping this picture in mind, we will study the groundstates for both Fredkin and Motzkin spin models in the continuum limit and construct possible effective field theory for them. In both spin models, the groundstates are in $S^z_{\rm tot}=0$ sector and have height representations which can be considered as the uniform superposition of Motzkin and Dyck paths, respectively.\cite{Sergey2012, DellAnna2016} 
The paths are defined in the upper half-plane and can be understood as the trajectories for a classical discrete random walk. In the continuum limit, the groundstate can be described in terms of coarse-grained height field $\phi$ with the constraint $\phi \geq 0$. 
The difference between Dyck and Motzkin paths lies in the diffusion constant of the random walk, carries over to the quantum wavefunction via a dimensionless 
parameter. 

We further construct a $z=2$ bosonic parent Hamiltonian for the continuum groundstate.
However, in contrast to the 2d quantum Lifshitz model, the $z=2$ bosonic field theory is not the effective theory for the Motzkin and Fredkin spin chains. 
Our careful numerical DMRG analysis of the dynamical exponent $z$ in the Motzkin \cite{short-prep} and Fredkin models suggests that $z$ lies close to 3
rather than 2: $z=3.2$. 
This large dynamical exponent in both spin models might be partially caused by the conservation of $S^z_{\rm tot}$.  
Similar behavior was found previously in the Kawasaki non-equilibrium dynamics for a classical Ising spin chain at low temperature, 
where the relaxation mode is governed by subdiffusive spin motion and has $z\simeq 3$.\cite{Kawasaki1966, Cordery1981,Grynberg}  

We further study the lowest excited state for the Fredkin model in the $S^z_{\rm tot}=0$ sector and find a different dynamical exponent $z_0=2.76$, indicating that this model has multiple dynamics. This behavior is common for example in metallic quantum critical points in higher dimensions,
where various order parameter fluctuations can have different dispersions.\cite{Hertz76,Millis93,ssbook,Oganesyan-2001,Meng2012,Lederer-2016} 

Finally, we study the stability of the Fredkin chain with respect to a ferromagnetic Heisenberg interaction; the strength of the
new interaction is proportional to the coupling $(1-\alpha)$, which varies from 0 to 1. 
At $\alpha=0, 1$, this model corresponds to Heisenberg and Fredkin models, respectively. 
We numerically explore the groundstate and dynamical exponents of our Fredkin-Heisenberg model. Away from the point $\alpha=1$, 
the analytical form of the groundstate is unknown. We use DMRG to study the groundstate and find that when $\alpha<1$, it is different from the
Fredkin case at $\alpha=1$, suggesting that the Fredkin model is unstable in the presence of the Heisenberg. 
Away from $\alpha=1$, the dynamical exponent for the lowest excitation drops to a value smaller than 3 and approaches 3 as we decrease $\alpha$ to zero. 
At $\alpha=0$, although the bulk is the ferromagnetic Heisenberg chain, the lowest excitation cannot be described by the 
$z=2$ diffusion mode due to the boundary condition being used, which favors up (down) spin on the left (right) boundary. 
Further study for the gapless low energy excitations in other spin sector demonstrates that the dynamical exponents can 
also be different in different spin sectors.

The structure of the paper is as follows. We first review both Fredkin and Motzkin spin models in Sec.~\ref{spin_model}. 
Then we discuss the ground in the continuum limit  and compute various local observables and entanglement properties in Sec.~\ref{orbifold_boson}.  
In Sec.~\ref{Fredkin_z}, we study the dynamical exponent for the Fredkin model, as well as its change in the presence of the Heisenberg interaction. 
We summarize and conclude in Sec.~\ref{conclusion}. In Appendix~\ref{Dyck_Motzkin}, we briefly explain some combinatorics used in Dyck and Motzkin path calculation. In Appendix~\ref{ap:dmrg}, we gives the detail for DMRG calculation in Fredkin-Heisenberg model.    

\section{Fredkin and Motzkin spin chains} 
\label{spin_model}

We introduce the Hamiltonians for the 2 quantum spin chains that are the focus of this paper. 
The first one is the Fredkin model \cite{DellAnna2016} and has
spin 1/2, while the second is the spin 1 Motzkin model \cite{Sergey2012}. We describe the exact groundstates of the 2 spin chains.

\subsection{Fredkin}
The spin $S=1/2$ Fredkin model \cite{DellAnna2016,Salberger2016} has the following Hamiltonian in the $S^z$ basis
\begin{align}
& H=H_{\rm bulk}+H_{\rm bdy}=\sum_{j=1}^{N-2} H_j+H_{\rm bdy},\nonumber\\
& H_j=(1+\sigma_j^z)(1-\vec{\sigma}_{j+1}\cdot \vec{\sigma}_{j+2})+ (1-\vec{\sigma}_j \cdot \vec{\sigma}_{j+1})(1-\sigma_{j+2}^z) \nonumber\\ 
& H_{\rm bdy}=|\!\downarrow\rangle_1 \langle \downarrow\!| + |\!\uparrow\rangle_N \langle\uparrow\!| 
\label{H_Fredkin}
\end{align} 
where $\vec \sigma=(\sigma^x,\sigma^y,\sigma^z)$ is the vector of Pauli matrices. We work with chains with an even number of sites $N$.
In the bulk Hamiltonian, $H_j$ can be expressed in terms of projectors: 
\begin{align}
  H_j=|\uparrow\rangle_j \langle\uparrow| \otimes |S\rangle_{j+1,j+2} \langle S|
  +|S\rangle_{j,j+1} \langle S|\otimes|\uparrow\rangle_{j+2} \langle\uparrow\! | 
\end{align} 
where $|S\rangle= \tfrac{1}{\sqrt 2} (|\!\uparrow\downarrow\rangle- |\!\downarrow\uparrow \rangle) $ is the singlet
built out of neighboring spins.   
We thus see that the bulk term $H_{\rm bulk}$ involves 3-spin interactions leading to two spin exchange processes:
$|\!\uparrow\uparrow\downarrow\rangle\Longleftrightarrow|\!\uparrow\downarrow\uparrow\rangle$ 
and  $|\!\downarrow\uparrow\downarrow\rangle\Longleftrightarrow|\!\uparrow\downarrow\downarrow\rangle$. These two moves are denoted as Fredkin gates in the field of quantum computation.  

\begin{figure}
\centering
\includegraphics[width=.42\textwidth]{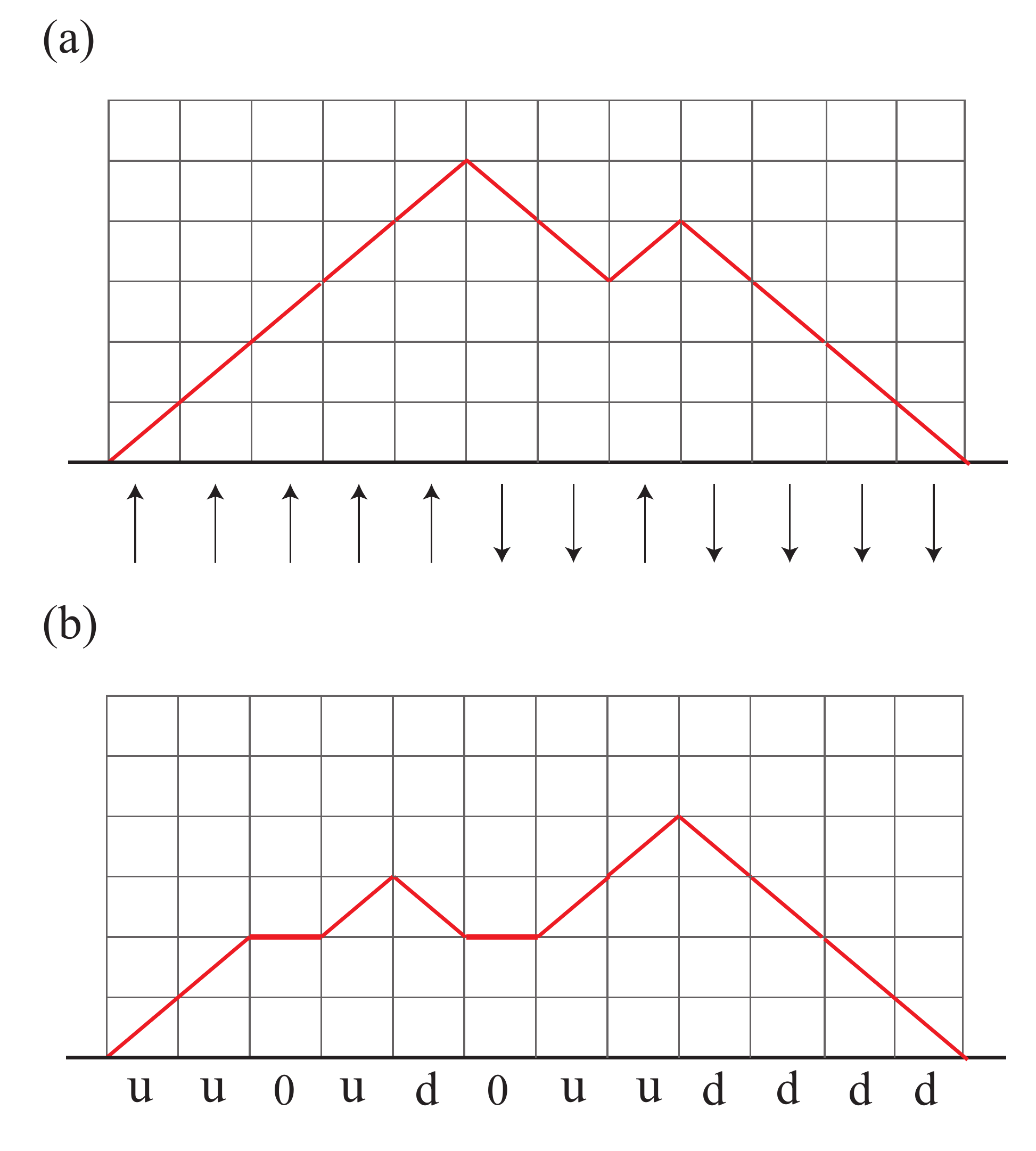}
\caption{a) A Dyck path and the corresponding spin configuration in the Fredkin model. The vertical spacing is $1/2$.
 b) A Motzkin path and the corresponding spin configuration in the Motzkin model. The vertical spacing is now $1$, which corresponds to the spin.}
\label{fig:Dyck_Motzkin}
\end{figure}

The Fredkin Hamiltonian is constructed in terms of projection operators which commute with the total   
$S^z_{\rm tot}=\sum_j S_j^z$ operator. The model therefore has $U(1)$ symmetry. 
The unique groundstate is known exactly and corresponds to an equal-weight superposition of states defined through \emph{Dyck paths}.\cite{DellAnna2016, Salberger2016} For example,
\begin{align}
  N=2:\quad |\Psi_0\rangle &= |\!\up\down\rangle = 
|\begin{tikzpicture}
\draw[gray, thick] (0,0) -- (.2,.2); 
\draw[gray, thick] (.24,.2) -- (.44,0);
\end{tikzpicture}\rangle \nonumber\\  
  N=4: \quad |\Psi_0\rangle &= \tfrac{1}{\sqrt 2} ( |\!\up\up\down\down\rangle + |\!\up\down\up\down\rangle ) = 
\tfrac{1}{\sqrt 2}\left( 
\left|\begin{tikzpicture}
\draw[gray, thick] (0,0) -- (.1,.1); 
\draw[gray, thick] (.12,.12) -- (.22,.22);
\draw[gray, thick] (.25,.22) -- (.35,.12);
\draw[gray, thick] (.37,.10) -- (.47,0);
\end{tikzpicture} \right\rangle
+|\begin{tikzpicture}
\draw[gray, thick] (0,0) -- (.2,.2); 
\draw[gray, thick] (.24,.2) -- (.44,0);
\draw[gray, thick] (.47,0) -- (.67,.2); 
\draw[gray, thick] (.69,.2) -- (.89,0);
\end{tikzpicture}\rangle
\right)
\label{eq:Dyck-GS}
\end{align}
A $N$-step Dyck path in the upper half-plane is naturally represented in terms of height variables $\phi_i$ defined through
\begin{align}  \label{height-rep}
  \phi_{i+1}-\phi_{i}= S_{i+1}^z
\end{align}
meaning that the spin is a discrete derivative of $\phi$: $S^z=\Delta \phi$. The height field is defined on the
``dual'' lattice so that its index runs from $0$ to $N$. 
In the height representation, since we have spin $S=1/2$, the state $|\!\uparrow\rangle$ maps to the step $(1, 1/2)$, 
$|\begin{tikzpicture}\draw[gray, thick] (0,0) -- (.2,.2);\end{tikzpicture}\rangle$, 
while $|\!\downarrow\rangle$ maps 
to the step $(1,-1/2)$,
$|\begin{tikzpicture}\draw[gray, thick] (0,.2) -- (.2,0);\end{tikzpicture}\rangle$.
The Dyck path is a succession of such steps; it starts at $(x,y)=(0,0)$, ends at $(x,y)=(N,0)$, and obeys the constraint that the height field $\phi_i$ does not
cross below the $x$-axis, i.e.\ $\phi_i\geq 0$. 
Examples are shown in Eq.\eqref{eq:Dyck-GS} and Fig.~\ref{fig:Dyck_Motzkin} (a).  
As stated below, the groundstate is an equal-weight superposition of all allowed Dyck paths and satisfies $H_j|GS\rangle=0$ and $H_{\rm bdy}|GS\rangle=0$. 
Such a groundstate is a generalization of the Rokhsar-Kivelson (RK) type wavefunction to one dimension. 
With some elementary combinatorics (Appendix~\ref{Dyck_Motzkin}), we can show that this groundstate has a large 
entanglement entropy for a bipartition into regions $A=[1,N_A]$ and $B=[N_A+1,N]$.
In the limit $N, N_A\to \infty$, the result is \cite{DellAnna2016}
\begin{align}
S_{\rm vN}=\frac{1}{2}\log\frac{N_A(N-N_A)}{N}+O(1)
\label{vN_EE}
\end{align}
When $N_A\ll N$, the EE scales as $\tfrac12\log N_A$, which takes a similar form as that for (1+1) dimensional CFTs with central charge $c=3$.\cite{calabrese_entanglement_2004} However, we will see that this model 
is not described by a CFT and is in fact less entangled.   
We mention in passing that the Fredkin model can be generalized to a half-integer spin model with $S>1$, where the groundstate is equal weight superposition of colored Dyck path and has a square-root violation of the area law.\cite{DellAnna2016,Salberger2016, Salberger2016b} This wavefunction can be further deformed into a weighted superposition of Dyck path with the groundstate properties and energy gap studied in  Ref.\ \onlinecite{Salberger2016b, Udagawa2017, Zhang_Klich_2017}. 

\subsection{Motzkin} 
Similarly, one can construct a spin $S=1$ RK type model with similar properties to the Fredkin chain.  
This model was actually introduced prior to the Fredkin one in Ref.~\onlinecite{Sergey2012}, and was further explored in 
Refs.~\cite{Movassagh_Shor_2016,Movassagh2017,short-prep}.  
The Hamiltonian is \cite{Sergey2012}  
\begin{align} \label{H_Motzkin}
H=\sum_{j=1}^{L-1}\Pi_{j,j+1}+H_{\rm bdy}
\end{align}
where $\Pi_{j,j+1}$ corresponds to nearest neighbor exchange processes and is defined in terms of projectors,
\begin{align}
\Pi_{j,j+1}\equiv |D\rangle_{j,j+1}\langle D|+|U\rangle_{j,j+1}\langle U|+|V\rangle_{j,j+1}\langle V|
\end{align} 
with
\begin{align}
|D\rangle=\frac{1}{\sqrt{2}}\left( |0d\rangle-|d0\rangle \right), \quad |U\rangle=\frac{1}{\sqrt{2}}\left( |0u\rangle-|u0\rangle \right)\quad|V\rangle=\frac{1}{\sqrt{2}}\left( |00\rangle-|ud\rangle \right)
\end{align}
$|u\rangle\equiv |+1\rangle, |d\rangle\equiv |-1\rangle$ and $|0\rangle$ are eigenstates of $S_j^z$.
The boundary term is the projector
\begin{align} \label{bdy-motzkin}
H_{\rm bdy}=|d\rangle_1\langle d|+|u\rangle_{N}\langle u|
\end{align}
On the left boundary, it favors both $|0\rangle$ and $|u\rangle$ states, while on the right boundary, it favors $|0\rangle$ and $|d\rangle$ states.
The bulk Hamiltonian is in fact of the anisotropic bilinear-biquadratic form:
\begin{align}
  \Pi_{j,j+1}= A_{ab}S_j^a S_{j+1}^b + B_{abcd}S_j^a S_j^b S_{j+1}^c S_{j+1}^d
\end{align}
where repeated spin indices are summed over. The coefficients $A,B$ are given in Ref.~\onlinecite{short-prep}.
The Motzkin Hamiltonian has a U(1) symmetry generated by $S_{\rm tot}^z$. \cite{Movassagh2017} 
Similarly to the Fredkin model, we can construct the groundstate of the Motzkin chain in the height representation 
with the identification $S^z=\Delta\phi$. Since the Motzkin model has $S=1$, $\phi$ can only jump by integers. 
The groundstate in the height representation \eqref{height-rep}  
is a uniform superposition of all paths connecting $(x,y)=(0,0)$ and $(x,y)=(N,0)$ in the  upper half-plane.  
These paths are formed by three types of moves: diagonal up $(1,1)$, $|\begin{tikzpicture}\draw[gray, thick] (0,0) -- (.2,.2);\end{tikzpicture}\rangle$, diagonal down $(1,-1)$, $|\begin{tikzpicture}\draw[gray, thick] (0,.2) -- (.2,0);\end{tikzpicture}\rangle$ and flat $(1,0)$, $|\begin{tikzpicture}\draw[gray, thick] (0,.1) -- (.2,.1);\end{tikzpicture}\rangle$ which correspond 
to the three states $|u\rangle$, $|d\rangle$ and $|0\rangle$ in the $S^z$ basis, respectively. Such a type of path is called 
a Motzkin path; one example is shown in Fig.~\ref{fig:Dyck_Motzkin} (b). For example, when $N=3$, the groundstate of Eq.\eqref{H_Motzkin} is 
\begin{align}
\label{motzkin}
   \ket{\mc M_3} = \tfrac{1}{\sqrt 4}( 
|\begin{tikzpicture}
\draw[gray, thick] (-.32,0) -- (-.04,0); 
\draw[gray, thick] (0,0) -- (.28,0); 
\draw[gray, thick] (.32,0) -- (.6,0);
\end{tikzpicture}
\rangle 
+
|\begin{tikzpicture}
\draw[gray, thick] (-.32,0) -- (-.04,0); 
\draw[gray, thick] (0,0) -- (.2,.2); 
\draw[gray, thick] (.24,.2) -- (.44,0);
\end{tikzpicture}
\rangle 
+
|\begin{tikzpicture}
\draw[gray, thick] (0,0) -- (.2,.2); 
\draw[gray, thick] (.24,.2) -- (.44,0);
\draw[gray, thick] (.48,0) -- (.76,0); 
\end{tikzpicture}
\rangle 
+
|\begin{tikzpicture}
\draw[gray, thick] (0,0) -- (.2,.2); 
\draw[gray, thick] (.24,.2) -- (.52,.2);
\draw[gray, thick] (.56,.2) -- (.76,0); 
\end{tikzpicture}
\rangle 
)
\end{align}
\label{motzkin} 

The Motzkin path wavefunction also has large EE and the leading term in EE is the same as that for Fredkin model shown in Eq.\eqref{vN_EE}. Similarly, there is also a higher integer spin model with $S>1$ with the groundstate as the equal weight superposition of colored Motzkin path.\cite{Movassagh_Shor_2016, Zhang_Klich_2016} 

In both Dyck path and Motzkin path wavefunctions, for each height configuration, we have the constraint $S^z_{\rm tot}=\phi_N-\phi_0=0$, which gives rise to the following groundstate expectation value \cite{Movassagh2017} 
  \begin{align}
    \langle S^\pm_i\rangle= 0
  \end{align}
where $S^\pm_i=S_i^x\pm i S_i^y$.
In contrast, $S^+_i S^-_j$ commutes with $S^z_{\rm tot}$, and we find that 
the two-point correlation function $\langle S^+_i S^-_j\rangle$ takes
a finite value, as we discuss in the next section. 

\section{Continuum wavefunction and a parent Hamiltonian}
\label{orbifold_boson}

Although the groundstate for the Fredkin and Motzkin lattice models are known exactly (as given above), it will prove
 convenient to write down the wavefunctions in the continuum. This will allow for a simple and physical derivation of many results, and will make manifest the emergent symmetries of the models.
In particular, we will see that \emph{spatial} conformal invariance emerges in the bulk of the chain.   

\subsection{Continuum version of the Motzkin and Fredkin groundstate}

In the continuum limit, the groundstate for both Motzkin and Fredkin models can be written as  \cite{short-prep}
\begin{empheq}{align}  \label{eq:gs-orbi}
  \Psi_0[\phi] = \frac{1}{\sqrt{Z}} e^{-\frac{\kappa}{2} \int dx (\partial_x\phi)^2 } \prod_x \theta(\phi(x)) 
\end{empheq} 
where the bosonic $\phi(x)$ field represents the coarse-grained height variable introduced above.  
In the lattice model, the height
is set to zero at either end of the chain, hence we impose the following Dirichlet boundary condition on the quantum field
\begin{align}
  \phi(0)=\phi(L)=0
\end{align}
The normalization factor $Z$ takes the form of a (0+1)-dimensional partition function:
\begin{align} \label{Z}
  Z = \int_{\phi(0)=\phi(L)=0}\!\!\mc D\phi(x)\; e^{- \kappa\int_0^L dx (\partial_x\phi)^2 } \prod_x\theta(\phi(x))
\end{align}
where $\theta(z)$ is the Heaviside function that enforces $\phi$ to be non-negative 
to match the constraint of the lattice Dyck/Motzkin paths that appear in the groundstate (see Fig.~\ref{fig:Dyck_Motzkin}). 

Eq.\eqref{eq:gs-orbi} gives a unified picture for the groundstate of both the Fredkin and Motzkin models. 
The difference between the two lattice wavefunctions is encoded in the parameter $\kappa$, and can be understood 
in terms of a random walk problem. The Dyck and Motzkin paths describe a one dimensional random walk on 
the non-negative half-integers and integers, respectively, with the horizontal axis of the path as the ``time'' direction, Fig.~\ref{fig:Dyck_Motzkin}. The random walks are constrained to start and finish at the origin $\phi=0$, which is called a Brownian
excursion. 
The wavefunction \eqref{eq:gs-orbi} is then probability of a given random path, and its specific form follows from the Legendre equation 
obeyed by $\phi(x)$ \cite{short-prep}.   
The diffusion constant of the random walk is $1/(4\kappa)$, and takes different values for the Dyck and Motzkin chains. 
For the Dyck type random walk (Fredkin model), the variance at a typical step is $\sigma^2=((1/2)^2+(-1/2)^2)/2=1/4$, 
while for the Motzkin type random walk, the variance at a typical step is $\sigma^2=(1^2+0^2+(-1)^2)/3=2/3$. Since the diffusion constant is given by $1/(4\kappa)=\sigma^2/2$, we have $\kappa=2, 3/4$ in the Fredkin and Motzkin groundstates, respectively.  

We can now compute various groundstate properties in the continuum limit. Let us start with the expectation value of $S^z$.
By virtue of Eq.\eqref{height-rep}, in the continuum limit we have 
\begin{align}
  S^z(x) = \partial_x\phi
\end{align}
We thus have $\langle S^z(x)\rangle=\partial_x\langle\phi(x)\rangle$. The expectation value of the height field is easily computed by 
mapping the calculation to that of an elementary quantum mechanics problem \cite{short-prep}, which offers yet a different perspective on the groundstate.    
We map the height variable to the position of the quantum particle, $\phi\to X$, and the position to imaginary time, $x\to\tau$. 
The expectation value then maps to Feynman path integral
\begin{align} \label{map_to_QM}
  \langle \Psi_0|\phi(x)|\Psi_0\rangle \to \frac{\int_{X(0)=X(L)=0}\mathcal D X\, X(\tau) e^{-S_1[X]} }{\int_{X(0)=X(L)=0}\mathcal D X\, e^{-S_1[X]} }
\end{align} 
with the constraint that the particle starts at the origin and ends there at time $L$.
$S_1$ is the Euclidean action of the particle:
\begin{align}  \label{QM-action}
  & S_1=\int_0^L d\tau \left[ \kappa \left(\frac{dX}{d\tau}\right)^2 + V(X)  \right] \\
 &V(X\!<\!0)=\infty, \; V =0 \mbox{ otherwise} \nonumber
\end{align}
where the potential $V$ is simply a hard wall that prevents the particle from penetrating the region $X<0$.
We can thus rewrite \eqref{map_to_QM} as 
\begin{align}
  \frac{\langle 0,L| \hat X(\tau) |0,0\rangle}{\langle 0,L| 0,0\rangle} 
  &= \int_0^\infty dX \frac{\langle 0,L|X,\tau\rangle X \langle X,\tau|0,0\rangle}{\langle 0,L|0,0\rangle} \\
  &= \int_0^\infty dX f(X,\tau) X
\end{align}
where we have used the eigenstates $|X,\tau\rangle$ of $\hat X(\tau)$ in the Heisenberg representation. $f$ is the probability distribution
of finding the particle at position $X$ at time $\tau$, and is thus normalized $\int_0^\infty dX f(X,\tau)=1$. 
We have restricted the
integrals to positive values of $X$ due to the potential. After evaluating the propagator $\langle X_f,\tau_f|X_i,\tau_i\rangle$ \cite{short-prep}, 
we find
\begin{align}  \label{eq:f-orb}
  f(\phi,x) = \frac{1}{2\sqrt{\pi}} \left(\frac{4\kappa L}{ (L-x)x}\right)^{\!3/2} \! \phi^2\exp\left[\frac{-\kappa L \phi^2}{ (L-x)x} \right] 
\end{align}
where we have reverted back to the field theory formulation in terms of $\phi,x$.
We thus get
\begin{align}
  \langle \phi(x)\rangle &= \int_0^\infty d\phi f(\phi,x) \phi = 2\sqrt{\frac{x (L-x)}{\pi \kappa L} } \label{phi-vev} \\
\langle S^z(x)\rangle &= \frac{1}{\sqrt{\pi\kappa L}} \frac{L-2x}{\sqrt{x(L-x)}} \label{sz-vev}                          
\end{align} 
As expected, we find that $\langle\phi\rangle$ vanishes at the boundaries and takes its maximal 
value at the middle point, $\sqrt{L/(\pi\kappa)}$ with the characteristic square root dependence of Brownian motion. 
The $S^z$ expectation value Eq.\eqref{sz-vev} goes from positive at $x<L/2$ to negative at $x>L/2$, which is a consequence of the boundary conditions
in the lattice models Eqs.(\ref{H_Fredkin},\ref{bdy-motzkin}) that favor the up (down) spin on the left (right) boundary. 
The expectation value rapidly approaches zero deep in the bulk.
Letting $x=\tfrac{L}{2}+a$ and considering the limit $L\gg a$, we find that $\langle S^z\rangle\sim a/L^{3/2}$, as shown in Table~\ref{table_EE}.  This result matches the calculation of Refs.\ \onlinecite{Movassagh2017, DellAnna2016} 
for the groundstate of the Motzkin model. 

\begin{figure}
\centering
\includegraphics[width=.45\textwidth]{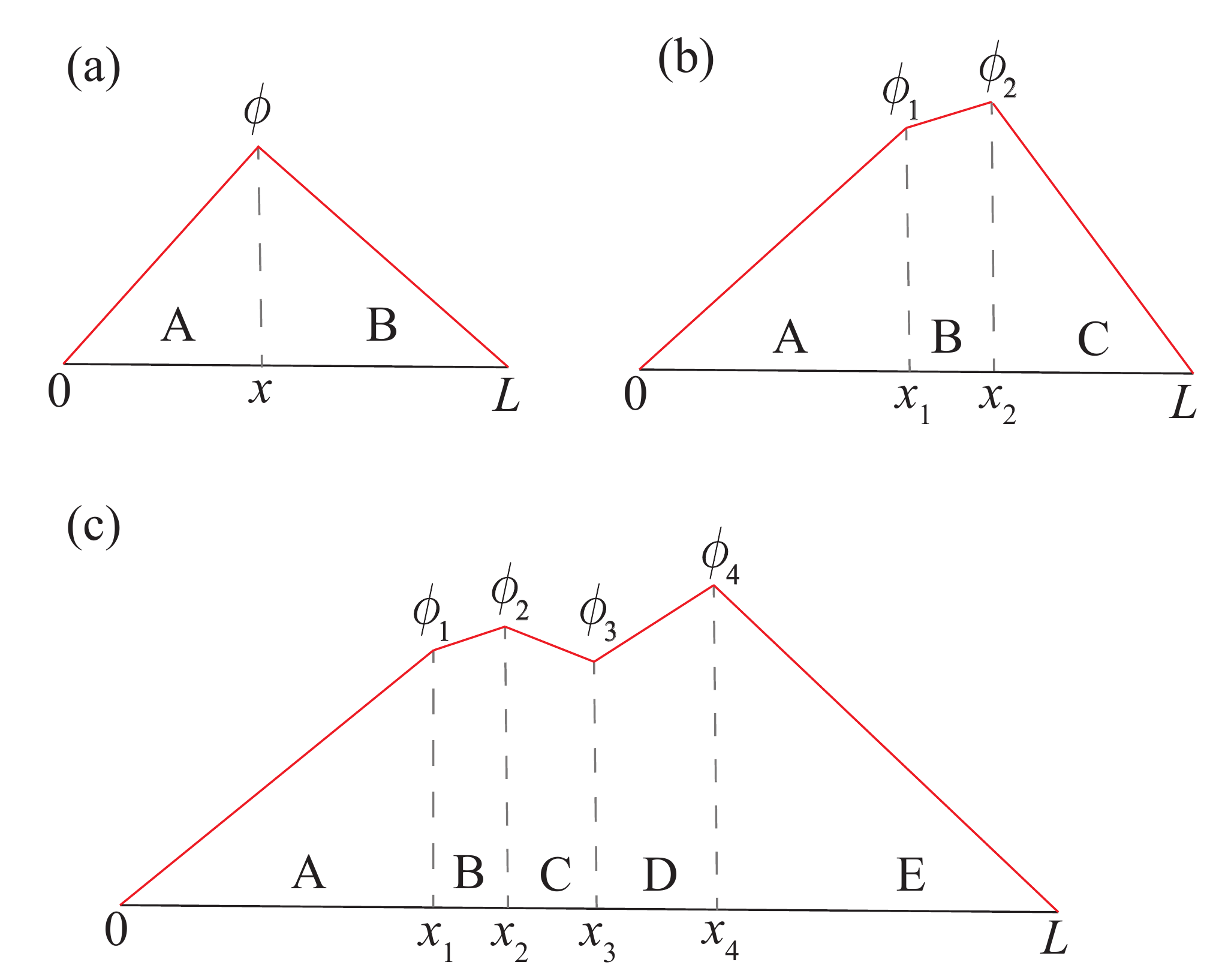}
\caption{(a) one configuration with $\phi$ at position $x$. (b) one configuration with $\phi_1$ and $\phi_2$ at position $x_1$ and $x_2$. (c) one configuration with $\phi_i$  at position $x_i$ with $1\leq i\leq 4$.}
\label{fig:corr_fun}
\end{figure}

Using a similar method, we can calculate the joint probability distribution function for a path to have its 
height equal to $\phi_1$ at $x_1$  
and $\phi_2$ at $x_2$ (Fig.~\ref{fig:corr_fun} (b)):
\begin{align}
f_2(\phi_1,x_1;\phi_2,x_2) 
&=\frac{1}{2\pi}\sqrt{\frac{\kappa}{x_2-x_1}}\left[ \frac{4\kappa L}{(L-x_2)x_1} \right]^{3/2}\phi_1\phi_2e^{-\frac{\kappa\phi_1^2}{x_1}-\frac{\kappa\phi_2^2}{(L-x_2)}}\left[ e^{-\frac{\kappa(\phi_1-\phi_2)^2}{(x_2-x_1)}}-e^{-\frac{\kappa(\phi_1+\phi_2)^2}{(x_2-x_1)}} \right]
\label{f_12}
\end{align}
Considering the limit deep inside the bulk with $0<x_2-x_1\ll x_1, L-x_2$, we find
\begin{align}
f_2 
\simeq \frac{1}{2\pi}\sqrt{\frac{\kappa}{x_2-x_1}}\left[ \frac{4\kappa L}{(L-x_2)x_1} \right]^{3/2}\phi_1(\phi_1+\phi_B)e^{-\frac{\kappa\phi_1^2}{x_1}-\frac{\kappa (\phi_1+\phi_B)^2}{L-x_2}} e^{-\frac{\kappa\phi_B^2}{x_2-x_1}}
\label{f_approx}
\end{align}
where $\phi_B=\phi_2-\phi_1$.  
It is easy to verify that Eq.\eqref{f_approx} is properly normalized: 
$\int_0^{\infty}d\phi_1\int_{-\infty}^{\infty}d\phi_B  f_2 =1$.
We can now obtain the height and $S^z$ auto-correlation functions in the groundstate:
\begin{align}
\langle\phi(x_1)\phi(x_2)\rangle  &= \int_0^{\infty}d\phi_1\int_{-\infty}^{\infty}d\phi_B\, \phi(x_1)\phi(x_2)f_2(\phi_1,x_1;\phi_2,x_2) = \frac{3x_1(L-x_2)}{2\kappa L}+ \cdots \\
\langle S^z(x_1) S^z(x_2)\rangle &= \langle\partial_{x_1}\phi(x_1)\partial_{x_2}\phi(x_2)\rangle= -\frac{3}{2\kappa L} + \cdots
\end{align}
where the dots denote subleading terms. {\color{black}pOnce $\kappa$ is set to the correct value (see above),} these results exactly match the calculation of Ref.~\cite{Movassagh2017} obtained using a different approach. 
We see that the $S^z$ correlator vanishes in the thermodynamic limit $L\to\infty$. 


\begin{table}
\centering
\begin{tabular}{c|c}
Groundstate & Properties deep  inside the bulk   \\\hline
$\langle S^z(\frac{L}{2}+ a ) \rangle$  & $-4\sqrt{\frac{1}{\kappa\pi}}\frac{a}{L^{3/2}}\to 0$ \\
$\langle S^z(x_1)S^z(x_2)\rangle$ & $-\frac{3}{2\kappa L}\to 0$  \\
$\langle S^+_iS^-_j\rangle$  & Fredkin: $\frac{1}{4}$, $\;$ Motzkin: $\frac{8}{9}$  \\
R\'enyi EE for an interval & $S_n(B)=\frac{1}{2}\log\frac{L_B}{\epsilon}+O(1)$  \\
Mutual information for 2 intervals & 0 \\
\end{tabular}
\caption{Summary of the groundstate properties of Eq.\eqref{eq:gs-orbi}, and its connection with the Motzkin and Fredkin models. These apply in the limit where
the length of the chain far exceeds all other scales. Also, $x_1\neq x_2$.
}\label{table_EE}
\end{table}

\begin{figure}
\centering
\includegraphics[width=.65\textwidth]{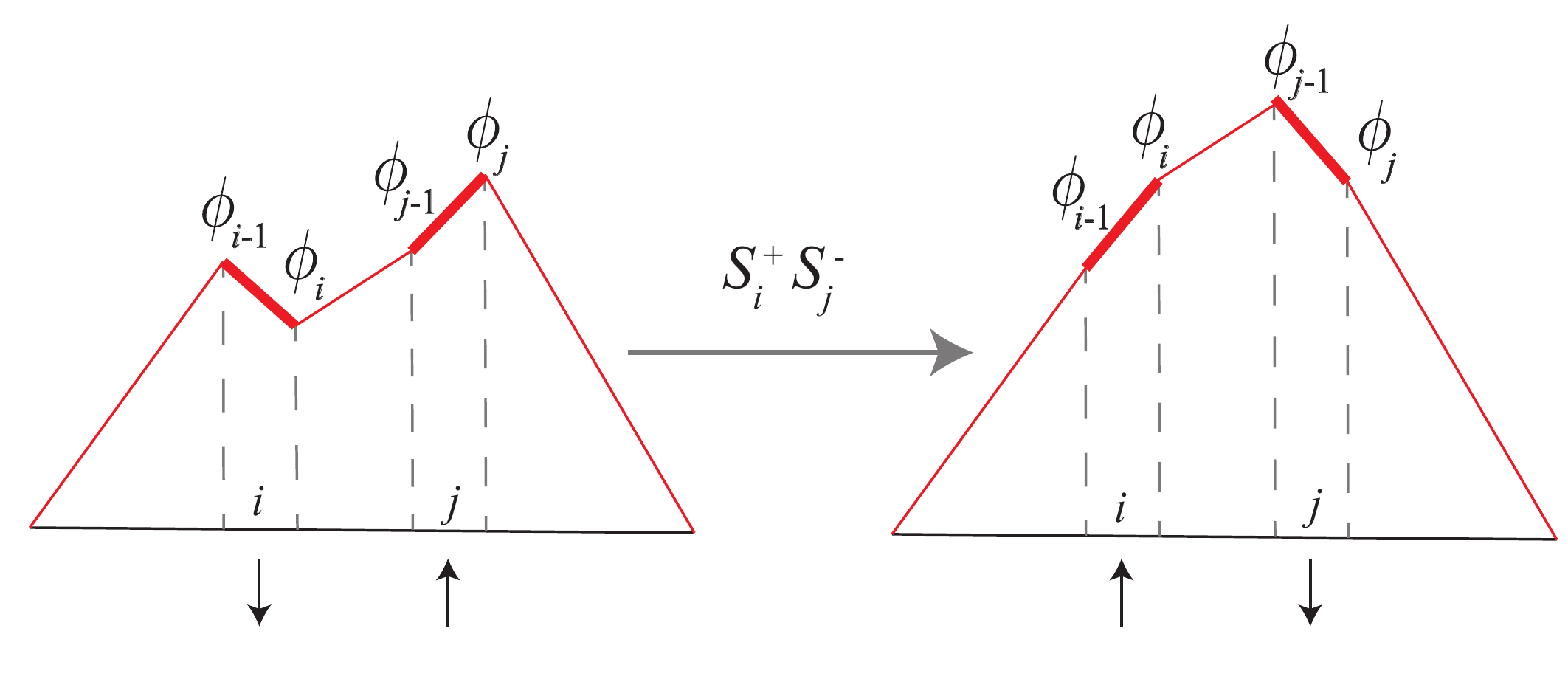}
\caption{Under the action of the operator $S^+_iS^-_j$, a height configuration (left) for a Dyck path with $S^z_i=-1/2$ and $S^z_j=1/2$ 
becomes the configuration (right) with $S^z_i=1/2$ and $S^z_j=-1/2$. In other words, the height field $\phi_\ell$ is shifted upwards by $2S=1$ along the
string $i\leq \ell <j $.
}  
\label{fig:S_iS_j}
\end{figure}

On the other hand, $\langle S^+_i S^-_j\rangle$ takes a non-zero value, and turns out to be related to a non-local string operator in the height representation. Let us consider this correlator in the lattice models.
In the Fredkin spin chain, when $S^+_i S^-_j$  acts on a height configuration $|h\rangle$, we have (assuming $i<j$)
\begin{align} 
S^+_i S^-_j|h\rangle= \begin{cases} |h^{\prime}\rangle, & \mbox{if } S^z_i|h\rangle=-\frac{1}{2}|h\rangle\ \mbox{and}\ S^z_j|h\rangle=\frac{1}{2}|h\rangle \\ 
0, & \mbox{if } S^z_i|h\rangle=\frac{1}{2}|h\rangle\ \mbox{or}\ S^z_j|h\rangle=-\frac{1}{2}|h\rangle\end{cases}
\end{align}
where $h$ and $h^{\prime}$ differ on the interval between $i$ and $j$, as shown in Fig.~\ref{fig:S_iS_j}. 
The action of $S^+_i S^-_j$ is thus to raise the height variable $\phi_\ell$ by 1 for $\ell$ belonging to the string $i\leq \ell < j$.
The effect of the shift along this string only changes the spins at sites $i$ and $j$. 
Now, since the groundstate is an equal weight superposition of all allowed height configurations (Dyck paths), 
$\langle S^+_i S^-_j\rangle$ is equivalent to $P(S^z_i=-1/2, S^z_j=1/2)$, namely the probability for the configuration with $S^z_i=-1/2$ and $S^z_j=1/2$. Deep inside the bulk, we have seen above that the two spins at $i$ and $j$ are uncorrelated, 
which implies that $P(S_i^z=-1/2,S_j^z= 1/2)=P(S_i^z=-1/2)P(S_j^z=1/2)=1/2\times 1/2=1/4$.

Starting from $\langle S^+_iS^-_j\rangle$ we can compute related correlation functions: deep in the bulk, we have
\begin{align}
\langle S^x_iS^x_j\rangle=\langle S^y_iS^y_j\rangle =\frac{1}{4}\langle S^+_iS^-_j + S^+_j S^-_i\rangle=\frac{1}{8}
\end{align}
and $\langle S^x_iS^y_j\rangle=0$. We can further generalize these results to the Motzkin model with $S=1$. The ladder operator satisfies
\begin{align}
S^+|d\rangle=\sqrt{2}|0\rangle,\quad S^+|0\rangle=\sqrt{2}|u\rangle,\quad S^+|u\rangle=0
\end{align}  
Therefore we have $\langle S^+_iS^-_j\rangle=\sqrt{2}\times 2/3\times\sqrt{2}\times 2/3=8/9$. 

\subsubsection{Entanglement}

We now explain the entanglement results listed in Table~\ref{table_EE}. 
Starting from $f_2$ defined in Eq.\eqref{f_approx}, we can calculate the reduced density for an interval $B$, $[x_1,x_2]$, 
that lies deep in the bulk (Fig.~\ref{fig:corr_fun}(b)). 
\begin{align}
\rho({\varphi}_B)=\frac{\sqrt{\epsilon}}{2 }\sqrt{\frac{4\kappa}{\pi (x_2-x_1)}}
\exp\left(-\frac{\kappa\epsilon{\varphi}_B^2}{x_2-x_1}\right) |{\varphi}_B\rangle\langle {\varphi}_B|
\end{align}
where 
\begin{align}
  \varphi_B= \frac{\phi_B}{\epsilon^{1/2}} = \frac{\phi_2 - \phi_1}{\epsilon^{1/2}}
\end{align}
is the dimensionless height difference between the endpoints of $B$ that also appeared in $f_2$, Eq.\eqref{f_approx}.  $\rho(\varphi_B)$ takes the form of 
a Gaussian distribution. 
The resulting R\'enyi EE is 
\begin{align}  \label{ee-simple}
  S_n(B)=\frac{1}{2}\log \left(\frac{L_B}{\epsilon}\right) + \log(\sqrt{\pi/\kappa})-\frac{1}{2(1-n)}\log n
\end{align}
where $L_B=x_2-x_1$. The leading term has a logarithmic dependence on the subsystem length,
which is the same as Eq.\eqref{vN_EE} for when the subsystem is the interval $[0, L_A]$. 
This logarithmic dependence on the length of the subsystem is also similar to that for $(1+1)$d CFTs.\cite{calabrese_entanglement_2004} 
However, the prefactor does not depend on the R\'enyi index, in contrast to what happens for CFTs. Interestingly, the same logarithm in the EE
was found in the XXZ spin chain, by taking a specific limit approaching the isotropic ferromagnetic interaction.\cite{Doyon11,Lauchli2012} 
Moreover, Eq.\eqref{ee-simple} is consistent with the EE result on an open cylinder for the (2+1)D version of the wavefunction \eqref{eq:gs-orbi} but with the $\phi\geq 0$ constraint removed.
There, in the thin torus limit the same leading logarithm was obtained.\cite{Xiao2017}  

Finally, we compute the mutual information between two disjoint intervals $B$ and $D$ deep inside the bulk,   
as shown in Fig.~\ref{fig:corr_fun}(c). The constraint $\phi>0$ can again be removed because the probability that $\phi$
approaches zero is vanishingly small deep in the bulk.
In this case, the joint distribution function for a path to have its height equal to $\phi_i$ at $x_i$, where $i$ runs from 1 to 4, is 
\begin{align}
f_4  = \frac{\kappa^2}{\pi^2}\sqrt{\frac{L}{L_AL_BL_CL_DL_E}} \exp\left(-\frac{\kappa\phi_1^2}{ L_A}-\frac{\kappa(\phi_2-\phi_1)^2}{ L_B}-\frac{\kappa(\phi_3-\phi_2)^2}{ L_C}-\frac{\kappa(\phi_4-\phi_3)^2}{L_D}-\frac{\kappa\phi_4^2}{ L_E}\right)
\label{f_two}
\end{align}  
where $L_\#$ denotes the length of interval $\#$. The reduced density matrix for the region 
$B\cup D$ only depends on $\phi_B\equiv\phi_2-\phi_1$ and $\phi_D\equiv\phi_4-\phi_3$ and can be obtained by integrating over $\phi_1$ and $\phi_2$ in $f_4$,
\begin{align} 
  \rho_{B\cup D}=\frac{\kappa\epsilon}{\pi}\sqrt{\frac{L}{(L-L_B-L_D)L_BL_D}}e^{-\frac{\kappa L_D(L-L_D)\epsilon{\varphi}_B^2+\kappa L_B(L-L_B)\epsilon{\varphi}_D^2 +2\kappa L_BL_D\epsilon{\varphi}_B{\varphi}_D}{L_BL_D(L-L_B-L_D)}}|{\varphi}_B,{\varphi}_D\rangle\langle {\varphi}_B,{\varphi}_D|
  \label{rho_BD}   
\end{align}
where $\varphi_B=\phi_B/\sqrt{\epsilon}$ and $\varphi_D=\phi_D/\sqrt{\epsilon}$. The R\'enyi entropy for $B\cup D$ is
\begin{align}
S_n(B\cup D)=\frac{1}{2}\log\left(\frac{L_BL_D}{\epsilon^2}\right)+\frac{1}{2}\log\frac{L-L_B-L_D}{L}+\log(\pi/\kappa)-\frac{1}{(1-n)}\log n
\end{align}
Similarly, we can also calculate EE for each interval and we have
\begin{align}
S_n(B/D)&=\frac{1}{2}\log(\pi/\kappa)+\frac{1}{2}\log \frac{L_{B/D}}{\epsilon}+\frac{1}{2}\log\frac{L-L_{B/D}}{L}-\frac{1}{2(1-n)}\log n
\end{align}
The mutual information between $B$ and $D$ is
\begin{align}
I_n(B,D)=S_n(B)+S_n(D)-S_n(B\cup D)=\frac{1}{2}\log\frac{(L-L_B)(L-L_D)}{(L-L_B-L_D)L}
\end{align}
$I_n(B,D)$ is independent of R\'enyi index and in the limit $L_B,L_D\ll L$, $I_n(B,D)$ tends to zero. 
This result demonstrates that this wavefunction is less entangled than generic $1+1$d CFTs, for which the mutual information scales as $1/r^{\Delta}$, where $r$ is the distance between two intervals and $\Delta$ is a function of the R\'enyi index and 
the scaling dimension of primary operators.\cite{Calabrese-2010}   

The Fredkin wavefunction, just as the Motzkin one \cite{short-prep}, provides an interesting example for which 
the EE is large but the mutual information is zero. The EE is a non-local quantity that detects entanglement between a subsystem 
and its complement, while the mutual information $I(A,B)=S(A)+S(B)-S(A\cup B)$, measures the correlations between the two intervals. 
The wavefunction Eq.\eqref{eq:gs-orbi} can be mapped to a constrained random walk problem. However, deep inside the bulk both the boundary conditions 
and $\phi \geq 0$ are not important and the random walk process can be described by the usual Brownian motion. The probability for the Brownian walker 
to move a distance $\delta \phi$ in ``time'' $\delta x$ is independent of the history. Therefore there is no correlation between two disjoint intervals and $I(A,B)=0$.
The mutual information also gives an upper bound for two-point correlation function of local observables.\cite{Wolf2008}  
This is consistent with the result  $\langle S^z(x_1)S^z(x_2)\rangle=O(L^{-1})\to 0$ for 
$x_1\neq x_2$. 

\subsubsection{Emergent conformal symmetry}
The analysis above has shown that deep inside the bulk, we can safely remove the $\phi \geq 0$ constraint and the wavefunction becomes,
\begin{align}  \label{Psi-no-constraint}
  \ket{\Psi_0} = \frac{1}{\sqrt{Z}}\int\mc D\phi(x)\; e^{-\frac{\kappa}{2} \int dx (\partial_x\phi)^2 } \ket{\phi(x)}
\end{align}
where $Z$ is the normalization factor:
\begin{align}  \label{Z-no-constraint}
  Z = \int\mc D\phi(x)\; e^{-\kappa \int dx (\partial_x\phi)^2 } 
\end{align}
Starting with the original wavefunction Eq.\eqref{eq:gs-orbi},  
we can expand the height field about its expectation value deep in the bulk, $\phi=\langle\phi\rangle+ \delta\phi$, with
$\langle\phi\rangle\sim \sqrt L$ as given in Eq.\eqref{phi-vev}. In the limit $L\to\infty$ for $x$ near the middle, the fluctuations
$\delta\phi$ thus become unconstrained and we recover Eq.\eqref{Psi-no-constraint}, which is free of the non-negativity constraint. 
Moreover, changing variables $x\to x-L/2$, we can effectively work on the infinite line to simplify the symmetry analysis.
The wavefunction \eqref{Psi-no-constraint} then enjoys an emergent $(1+0)$-dimensional conformal symmetry   
in the sense of conformal quantum mechanics.\cite{Jackiw1972, deAlfaro1976} 
We indeed recognize Eq.\eqref{Z-no-constraint} as the partition function of a free quantum mechanical particle.
It is straightforward to verify that the wavefunction is invariant under SL$(2,\mathbb R)$ M{\"o}bius transformations: 
\begin{align}
  x' &= \frac{\alpha x +\beta}{\gamma x +\delta} \\
  \phi'(x') &=\frac{\phi(x)}{\gamma x+\delta}
\end{align}
with the unit determinant condition: $\alpha\delta-\beta\gamma=1$. We emphasize that we are working deep in the bulk so that we can neglect total derivatives 
in the integrals. {\color{black} In the presence of boundaries, the above transformations
are no longer symmetries because even the translational symmetry is broken.}  
The SL$(2,\mathbb R)$ symmetry group has 3 real parameters:
it describes translations, dilations, and less obviously,
\emph{special conformal transformations}. The latter corresponds to $\alpha=\delta=1$ and $\beta=0$.  
The wavefunction can thus be viewed as a lower dimensional version of the wavefunctions of conformal quantum critical points 
introduced in (2+1)d, which have groundstates invariant under the infinite group of two-dimensional conformal transformations.\cite{Ardonne-2004,Fradkin-book}
As we shall see in the next section, Eq.\eqref{Psi-no-constraint} is the groundstate of the 1+1D version of the quantum Lifshitz model, but with a non-compact height field.

\subsection{Continuum parent Hamiltonian}   
We construct a Hamiltonian in the continuum limit which has the same groundstate as 
the Fredkin and Motzkin spin chains (naturally, also in the continuum limit). 
We consider the following quantum field theory (QFT) for the height field $\phi$,
\begin{align}
  \mc Z &=\int_{\phi(0,t)=\phi(L,t)=0} \mc D\phi(x,t) \, e^{iS[\phi]} \prod_{x,t}\theta(\phi(x,t)) \nonumber \\ 
  S&= \frac12 \int dt dx \, \big( -(\partial_t\phi)^2 + \kappa^2 (\partial_x^2\phi)^2 \big)
\label{height_action}
\end{align}
The partition function $\mathcal Z$ should not be confused with the normalization of the groundstate Eq.\eqref{Z}.  
$\theta(z)$ is the Heaviside function that enforces the field $\phi$ to be non-negative, to match the constraint 
of the lattice Dyck and Motzkin paths (see Fig.~\ref{fig:Dyck_Motzkin}).  
The action \eqref{height_action} describes a non-compact boson with the constraint that $\phi$ is non-negative. 
This can be interpreted in terms of an \emph{orbifold}.
Indeed, starting from the non-compact boson 
with target space $\mathbb R$, we mod out by the discrete $\phi$-parity symmetry of the unconstrained theory, $\phi\to -\phi$, 
which forms a $\mathbb Z_2$ group. This results in $\phi$ taking values in the 
orbifold $\mathbb R/\mathbb Z_2=[0,\infty)$, i.e.\ the semi-infinite real line.\footnote{The point $\phi=0$ is special because it is invariant under the $\phi\to-\phi$
transformation, and leads to $\mathbb R/\mathbb Z_2$ being an orbifold not a manifold. To get the latter, one would need to remove $\phi=0$.}
Alternatively, we can view the field $\phi$ as taking values in $\mathbb R$, but with a potential term $V(\phi)=\infty$ for 
$\phi<0$, and $V=0$ for $\phi \geq 0$ (see Eq.\eqref{QM-action}).     

Without the $\phi\geq 0$ constraint, the above QFT is the 1+1D version of the quantum Lifshitz model \cite{Ardonne-2004}, albeit with a non-compact height field. We shall adapt the methods previously used to study that theory in order to determine the low lying excitations.
We can write the Hamiltonian corresponding to Eq.\eqref{height_action} as 
\begin{align} \label{H-orbi}
  H &= \int dx \left( \frac12 \Pi^2 + \frac{\kappa^2}{2} (\partial_x^2\phi)^2 + V(\phi) \right) 
\end{align}
with the canonical equal-time commutation relation,
\begin{align}
  [\phi(x),\Pi(x')] = i\delta(x-x')
\end{align}
Such a Hamiltonian with general potential $V(\phi)$ was considered in \cite{Dijkgraaf}, where the 
relation between the $(d+1)$-dimensional quantum theory and its Euclidean $d$-dimensional ``seed'' was called
stochastic quantization. This nomenclature shall become clearer below when we discuss the mapping to non-equilibrium
dynamics of the classical spin chain.

It shall be useful to adopt the Schr\"odinger functional formalism, where $\Pi=-i\frac{\delta}{\delta\phi}$.
The Schr\"odinger equation for the eigenstates of $H$ then reads:
\begin{align}
  \int dx \left(-\frac12 \left( \frac{\delta}{\delta\phi}\right)^2 +\frac{\kappa^2}{2} (\partial_x^2\phi)^2 + V(\phi) \right) \Psi[\phi] = E \Psi[\phi] 
\end{align}
We then introduce the annihilation and creation operators $Q,Q^\dag$ \cite{Ardonne-2004,Fradkin-book}: 
\begin{align}
  Q(x)&=\frac{1}{\sqrt 2}\left( i\Pi -\kappa\partial_x^2\phi \right) 
  = \frac{1}{\sqrt 2}\left( \frac{\delta}{\delta\phi} -\kappa\partial_x^2\phi \right) \\
Q^\dag(x) &=\frac{1}{\sqrt 2}\left( -i\Pi -\kappa \partial_x^2\phi \right) 
  = \frac{1}{\sqrt 2}\left( -\frac{\delta}{\delta\phi} -\kappa \partial_x^2\phi \right)
\end{align}
Then we can express the Hamiltonian as 
\begin{align}
  H=\int dx \left(\frac12 \{Q^\dag(x), Q(x)\}+V(\phi)\right) - E_{\rm vac}  
  = \int dx \Big(Q^\dag(x) Q(x) + V(\phi)\Big)  
\end{align}
where we have subtracted the infinite groundstate energy, $E_{\rm vac}\!=\!-\int dx \partial_x^2\delta(x-y)|_{y\to x}\!>\! 0$ and the Hamiltonian is normal-ordered. The potential 
simply enforces $\phi \geq 0$. The groundstate has zero energy $E_0=0$ and is annihilated by all the $Q(x)$ operators: 
$Q\Psi_0[\phi] = 0$.
The corresponding linear functional differential equation, $(\frac{\delta}{\delta\phi} -\kappa\partial_x^2\phi)\Psi_0[\phi]=0$, has the unique solution Eq.\eqref{eq:gs-orbi}.  
To obtain this result we have used $\frac{\delta\Psi_0}{\delta\phi} = (\kappa\partial_x^2\phi )\,\Psi_0$.
In calculating this functional derivative, we have assumed that $\phi>0$ for all $x$, 
i.e.\ we have ``regularized'' the singular point $\phi=0$
of the orbifold by setting the smallest value of $\phi(x)$ to be $0^+$. This in particular ensures that the functional derivative
of the Heaviside function $\theta(\phi)$ vanishes. 

To get the excited states of Eq.\eqref{H-orbi}, we can act with the creation operator $Q^\dag$ on the groundstate   
\begin{empheq}{align}  \label{eq:exci-orbi}
  \Psi_{k}[\phi] &=  \int dx\, e^{i k x} Q^\dag(x) \Psi_0[\phi] \\
  H\Psi_{k}[\phi] &=\kappa k^2\, \Psi_{k}[\phi]  
\end{empheq}  
where we have omitted a normalization constant.
To obtain the energy eigenvalue, $\kappa k^2$, we have used:
\begin{align}
  [Q(x),Q^\dag(y)]= -\kappa\, \partial_x^2 \delta(x-y) 
\end{align}   
We thus see that the spectrum is gapless, with the low lying excitations having a $\kappa k^2$ dispersion, which implies that $z=2$. 
This is also manifest from the action \eqref{height_action}.

Going back to the lattice, previous exact diagonalization \cite{Sergey2012} and DMRG \cite{DellAnna2016} results suggest that $z>2$ for both the Fredkin 
and Motzkin models. In Sec.~\ref{Fredkin_z},
we conclusively show that the dynamics are indeed subdiffusive.
Therefore, Eq.\eqref{height_action} is not the correct effective field theory for the quantum spin models under consideration. It thus provides 
an example where a given state is the groundstate of Hamiltonians with qualitatively distinct excitation spectra. {\color{black} Moreover, it would be interesting to construct the correct effective theory for the Fredkin and 
Motzkin models to reflect the subdiffusive dynamics. We leave this task for future study.}  

\subsubsection{Liouville quantum mechanics}

As we discussed in Sec.~\ref{orbifold_boson}, the 1d wavefunction in Eq.\eqref{eq:gs-orbi} gives a path integral 
representation for a quantum mechanical particle restricted to move on the positive axis.  We can soften the hard wall constraint, and consider the following  wavefunction,
\begin{align}  
  \Psi[\phi] = \frac{1}{\sqrt{Z}} \exp\left({-\frac{1}{2}\int dx \left[ \kappa \left(\frac{d\phi}{d x}\right)^2  +\mu e^{-\lambda \phi}  \right] }\right)
\label{Liou_wf}
\end{align} 
with the parameters $\mu,\lambda>0$.  
As $\lambda\to \infty$, we recover the $\phi \geq 0$ constraint and when $\lambda\to 0$, 
the constraint disappears and the wavefunction reduces to the groundstate of the ordinary $z=2$ (non-compact) boson. 

Eq.\eqref{Liou_wf} describes the action for a particle moving in an exponential potential which is called Liouville quantum mechanics.\cite{DHoker1982} This originates from the zero mode part of Liouville field theory in $(1+1)$D and has been extensively studied in the context of string theory.\cite{POLYAKOV1981} More recently, Liouville field theory has been used to study Anderson localization in 2d; it gives the effective theory for 
the critical wavefunction after disorder averaging.\cite{Kogan1996} 
For the state in Eq.\eqref{Liou_wf}, the correlators can be evaluated as the quantum expectation  
values for Liouville quantum mechanics in imaginary time.\cite{Shelton1998, Balents1997}  

Starting from this wavefunction, we can define an annihilation operator $Q$,
\begin{align}
  Q= \frac{1}{\sqrt 2}\left( \frac{\delta}{\delta\phi} -\kappa\partial_x^2\phi-\frac{\mu \lambda}{2} e^{- \lambda \phi} \right) 
\end{align}  
It satisfies $Q|\Psi\rangle=0$. The $(1+1)$d RK Hamiltonian takes the form $\int dx Q^{\dag}Q$ 
with $\Psi[\phi]$ as the zero energy groundstate. It would be interesting to find a lattice spin model which can be described by this field theory.
Also, since there is an emergent conformal symmetry deep inside the bulk in both limits $\lambda=0,\infty$, it would be interesting 
to investigate what happens at intermediate couplings.

\section{DMRG analysis}  
\label{DMRG}  
In this section, we describe our large-scale DMRG calculations performed using the ITensor library for both the Fredkin 
and Motzkin models. We begin with the groundstate properties, and then move on to the excited states.

\begin{figure}[hbt]
\centering
 \subfigure[]{\label{fig:Sz} \includegraphics[width=.45\textwidth]{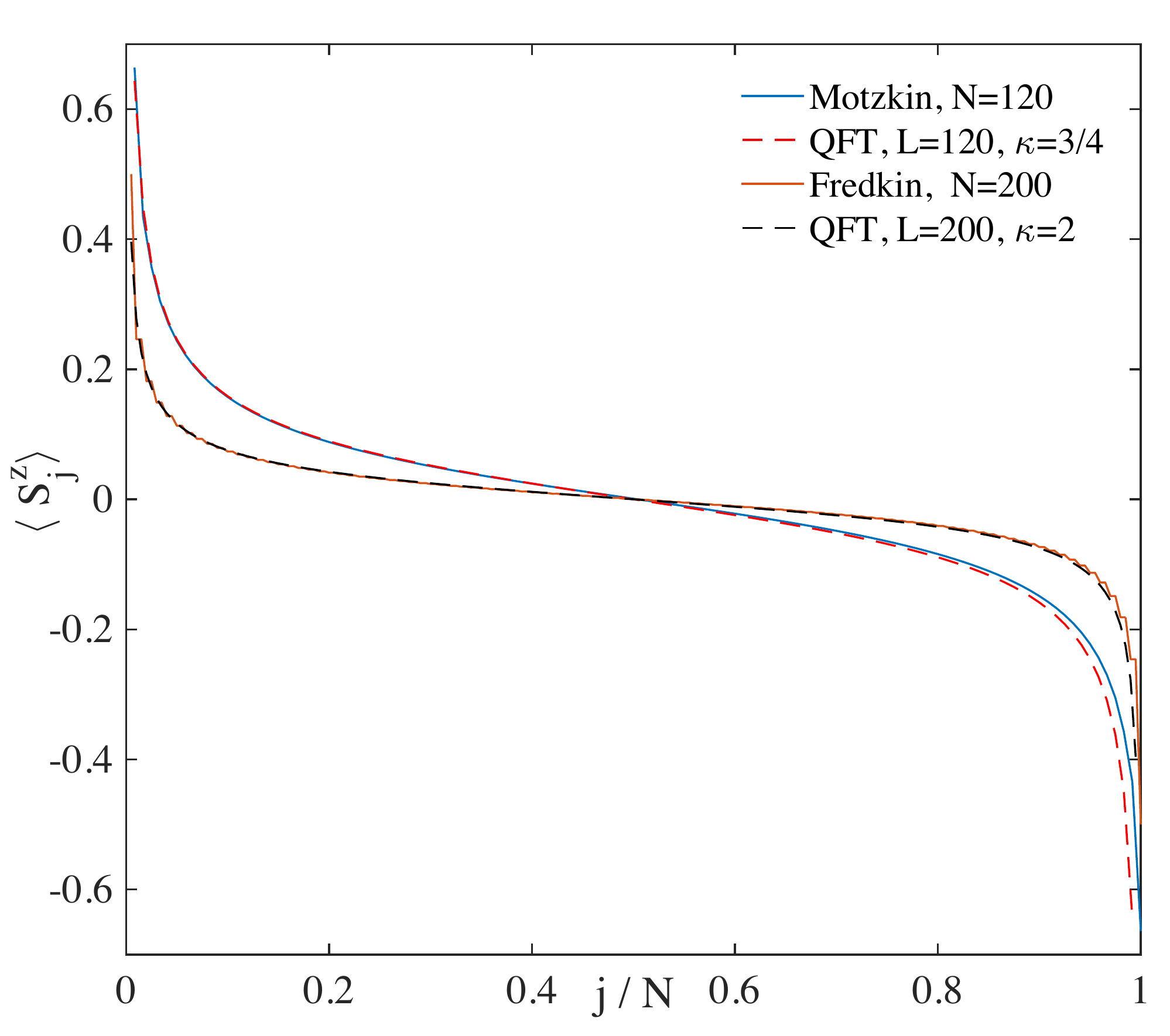}}
 \subfigure[]{\label{fig:SzSz} \includegraphics[width=.47\textwidth]{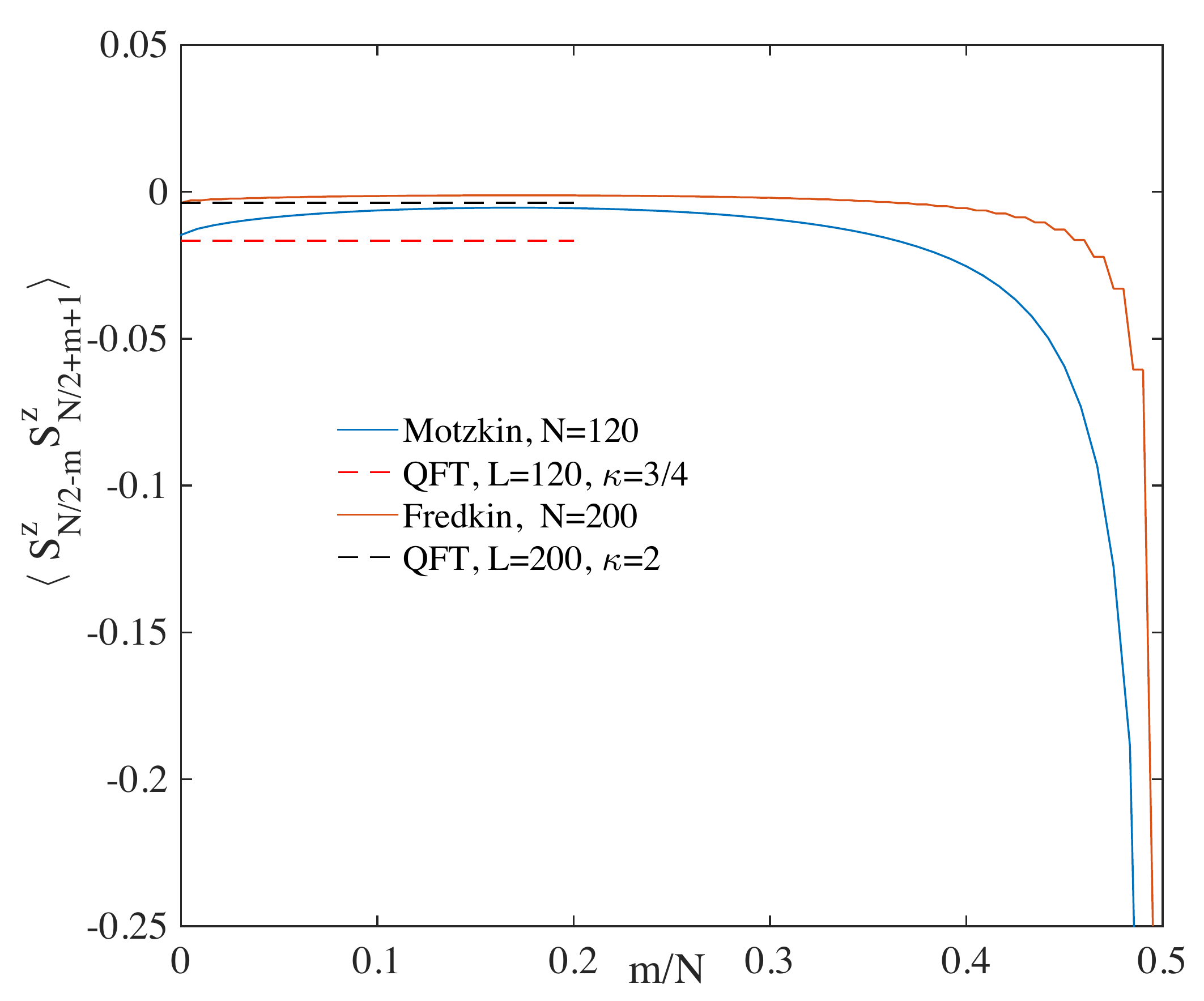}}   
\caption{
(a) $\langle S^z_j\rangle$ in both the Motzkin and Fredkin models obtained using DMRG (solid lines).  The dashed lines 
are the analytical field theory results. (b) $\langle S^z_{N/2-m}S^z_{N/2+m+1}\rangle$ in both Motzkin and Fredkin models obtained using DMRG. 
The dashed lines are the analytical field theory results deep in the bulk, which approach zero as $L\to\infty$.
}  
\label{fig:SzSz_corr}
\end{figure}

We first compute various correlation functions for the groundstate and compare them with the analytical results shown in Table \ref{table_EE}. Fig.~\ref{fig:Sz} shows $\langle S^z_j\rangle$  for both Motzkin and Fredkin models, which is close to zero in the bulk and takes nonzero values
near the boundaries. In the vicinity of the middle point, $\langle S^z_j\rangle$ deviates from zero linearly as a function of $j-N/2$.  
This is consistent with the analytical calculation in the continuum shown in Table~\ref{table_EE}. $\langle S^z_j\rangle$ approaches 
zero in the thermodynamic limit $L\to\infty$ for $j$ in the bulk, as shown in Eq.\eqref{sz-vev}.  
In Fig.~\ref{fig:SzSz}, we show $\langle S^z_{N/2-m}S^z_{N/2+m+1}\rangle$  for both Motzkin and Fredkin models.   
When $m$ is small, the correlator scales as $1/L$, which agrees with the analytical result deep inside the bulk. When $m\to N/2$, it 
becomes negative due to the boundary terms that favor anti-alignment between the left and right boundaries. 

\begin{figure}[hbt]
\centering
 \subfigure[]{\label{fig:Mot_corr} \includegraphics[width=.47\textwidth]{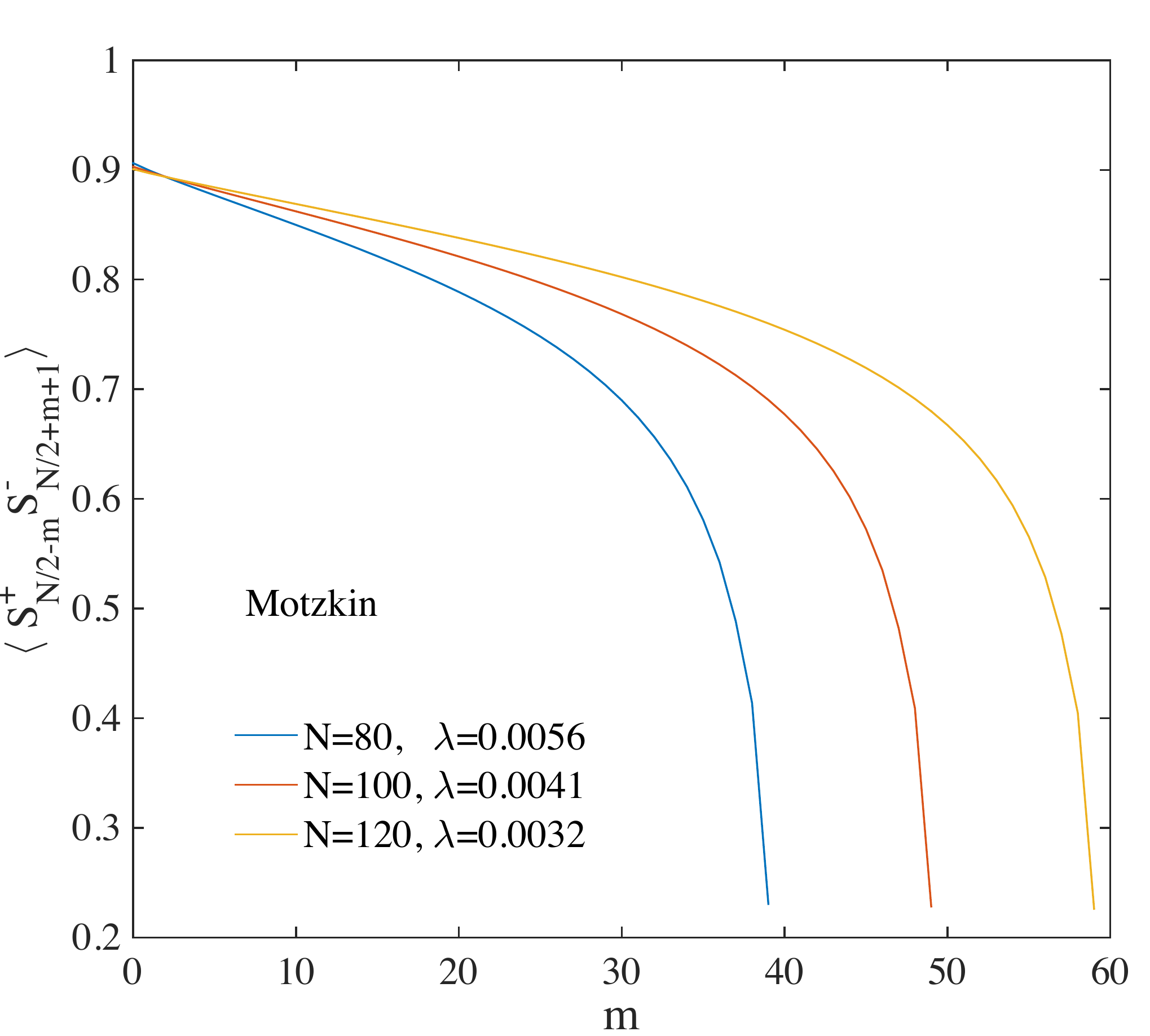}}
 \subfigure[]{\label{fig:Fre_corr} \includegraphics[width=.48\textwidth]{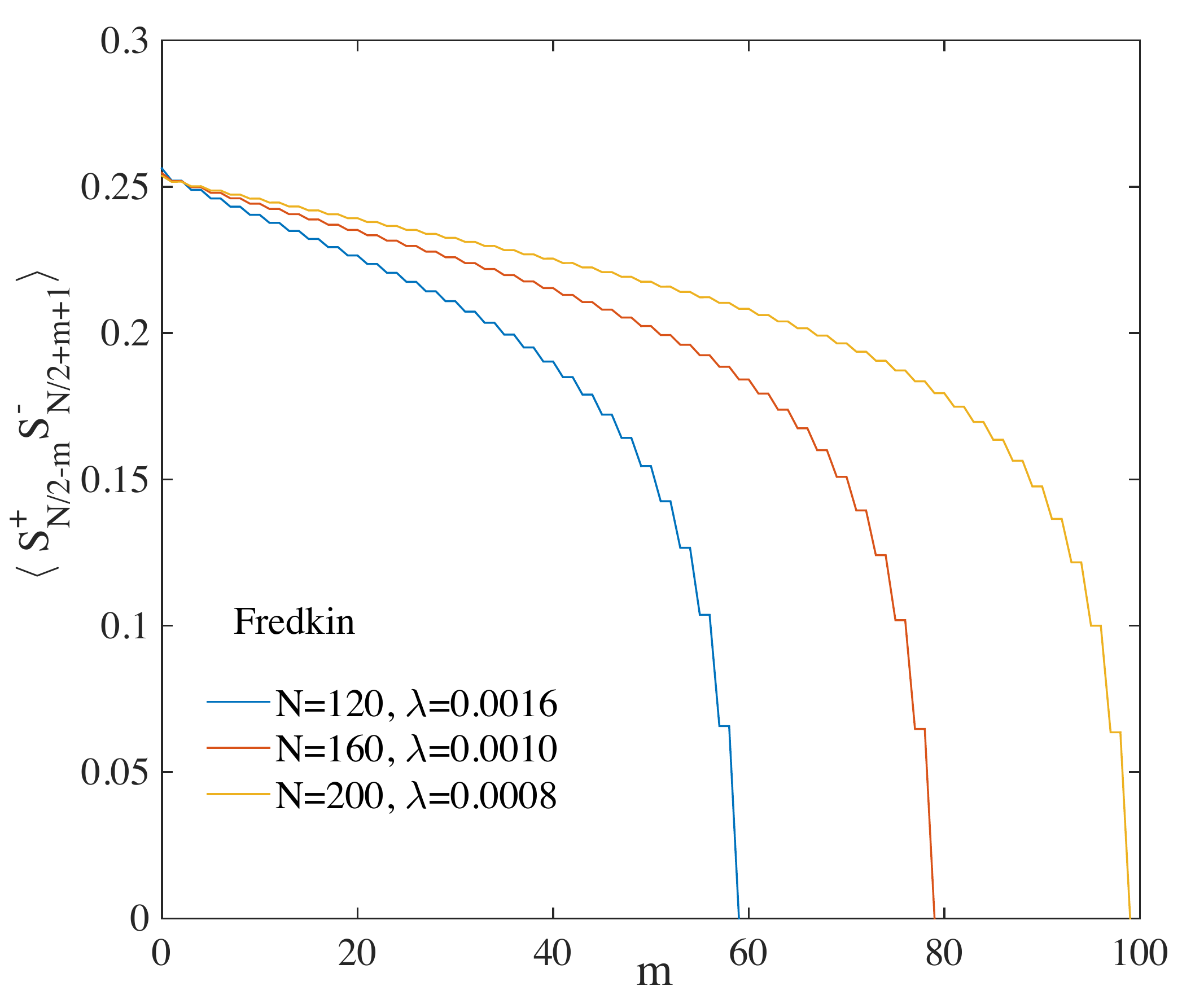}}   
\caption{
(a) The two-point correlation function $\langle S^+_{N/2-m} S^-_{N/2+m+1}\rangle$ in the Motzkin model for various system sizes. 
$\langle S^+_{N/2-m} S^-_{N/2+m+1}\rangle\sim 8/9-\lambda m$, where $\lambda$ decreases as $N$ increases. 
(b) The two-point correlation function $\langle S^+_{N/2-m} S^-_{N/2+m+1}\rangle$ in the Fredkin model for various system sizes. $\langle S^+_{N/2-m} S^-_{N/2+m+1}\rangle\sim 1/4-\lambda m$, where $\lambda$ decreases as $N$ increases.} 
\label{fig:Mot_Fre_corr}
\end{figure}

We also compute $\langle S^+_{N/2-m} S^-_{N/2+m+1}\rangle$, which is shown in Fig.~\ref{fig:Mot_Fre_corr}.   
In comparison with the $S^z$ auto-correlation function, we see that $\langle S^+S^-\rangle$ is much bigger when the separation is small compared
to the size. At small $m$, we find good agreement with the field theory predictions: $3/4$ (Fredkin), $8/9\approx 0.89$ (Motzkin). 
We notice that $\langle S^+_{N/2-m} S^-_{N/2+m+1}\rangle$ is not a constant but deviates from $1/4$ linearly as a function of $m$, i.e., $\langle S^+_{N/2-m} S^-_{N/2+m+1}\rangle=1/4-\lambda m$, which is a consequence of finite size. A similar situation happens for $\langle S^z(L/2+a)\rangle\sim a$, 
as shown in Fig.~\ref{fig:Sz}. 
We see that $\lambda$ is a function of $N$ that approaches zero as $N$ increases.  

\subsection{Dynamical exponents of the Fredkin model}  
\label{Fredkin_z} 
For the Fredkin model, the groundstate is unique and has $S^z_{\rm tot}=0$. We find that the lowest excited state has $S^z_{\rm tot}=\pm 1$, 
being doubly degenerate. This is confirmed both by exact diagonalization on small systems $N\leq 10$, and large-scale DMRG with $N$ up to $200$.
The energy gap scales as $1/N^z$ for sufficiently large $N$, where $z$ is the dynamical exponent of the quantum system.  
Since $z$ is large, we need to use large bond dimensions and enough sweeps in the DMRG computation to ensure 
convergence of the energy. We compute the gap $\Delta E$ for various system sizes, and fit the data to $1/N^z$, as shown in 
Fig.~\ref{fig:Fre_corr}.      
We find $z=3.23$, which is larger than $z=2.9$ obtained previously via DRMG in Ref.~\onlinecite{DellAnna2016}.  
Our result is consistent with the analytical bound obtained by Movassagh: $2 \leq z<13/2$.\cite{Movassagh_gap_2016}   
In addition, we have found a singly-degenerate excitation with $S_{\rm tot}^z=0$; its gap to the groundstate scales as $\Delta E_0\sim 1/N^{z_0}$,   
with a different dynamical exponent $z_0=2.76$, which is closer to the exponent of Ref.~\onlinecite{DellAnna2016}, suggesting that these authors worked in 
the $S^z=0$ symmetry sector, which does not contain the lowest excitations. These different dynamical exponents $z$ and $z_0$ for excitations in different sectors show that 
the model hosts multiple dynamics. 
These results are in close correspondence with our findings for the Motzkin model \cite{short-prep}, where we have found
$z=3.16$ and $z_0=2.71$, which are close to the Fredkin exponents.  
Our results for both models demonstrate that the $z=2$ boson theory presented above in Eq.\eqref{height_action} cannot 
be the effective field theory for the Fredkin and Motzkin Hamiltonians. It does describe their groundstates correctly, but not the excitations. 

We further analyze study the role of finite size effects on $z$ in Appendix \ref{ap:dmrg} by calculating $z((N_1+N_2)/2)$ from 2 consecutive values of $N$.   
This method was used in our previous paper for the Motzkin model \cite{short-prep}. 
For sufficiently large systems, we find that the variation of $z(N)$ with $N$ is small and roughly scales as $1/N$. 
We use a $1/N$ extrapolation to estimate $z(N\to\infty)$ and we find $z(\infty)=3.17$, in 
good agreement with the value given above. We also find $z_0(\infty)=2.69$.

\subsubsection{Mapping to non-equilibrium dynamics} 

We here point out a connection between the quantum dynamics of the chain and the \emph{non-equilibrium} relaxation of the classical spin chain \cite{short-prep}.
This provides hints regarding the large value of $z$. For a RK-type Hamiltonian like the Fredkin or Motzkin model, 
the quantum dynamics can be exactly mapped to the non-equilibrium dynamics of the corresponding classical spin chain.
The temporal evolution of the classical system is governed by a 
Markovian master equation for the probabilities $P_C(t)$ of the classical spin configurations $C$ \cite{Henley04} 
\begin{align}
  \frac{d P_C(t)}{dt} = \sum_{C'} W_{C,C'} P_{C'}(t) 
\end{align}
The rate matrix $W$ is related to the quantum Hamiltonian as follows:
\begin{align}
  W_{C,C'}= - \langle C|H|C'\rangle\,, \quad C\neq C'
\end{align}
where $C=\{S_i^z\}$ denotes a spin configuration in the $S^z$-basis.
The diagonal elements of $W$ are defined in order to satisfy detailed balance, $W_{C,C}=-\sum_{C', C'\neq C}W_{C',C}$. 
Under this mapping, the excited states of $H$ map to the classical relaxational modes of the rate matrix $W$.  
For example in the classical 1d Ising spin chain, endowed with Glauber dynamics, the relaxation can be described by the random walk of  
a single spin or a domain wall and leads to dynamical exponent $z=2$.\cite{Cordery1981} 
However, in our case we have a conserved U(1) symmetry generated by $S_{\rm tot}^z$, which maps to Kawasaki-type dynamics in the classical spin chain.
The constraint caused by the conservation law can effectively slow down the motion of the spins, and lead to subdiffusive dynamics $z>2$. 
In the Fredkin model, the corresponding classical dynamics are determined by these two kinds of moves: $\uparrow\uparrow\downarrow\Longleftrightarrow\uparrow\downarrow\uparrow$ and $\downarrow\uparrow\downarrow\Longleftrightarrow\uparrow\downarrow\downarrow$. A pair of adjacent opposite spins can be flipped if the third one is pointing in some special direction. This constraint slows down the motion of domains
leading to a larger $z$. 
A similar observation was made for the Motzkin model.\cite{short-prep}
It would be interesting to make this argument more precise in the future.    


\subsection{Crossover from Fredkin to Heisenberg} 

We perturb the Fredkin model with a pure Heisenberg ferromagnetic interaction in order to assess its stability,
which leads to the Fredkin-Heisenberg model:
\begin{align}
H_{\rm bulk}=\alpha H_F+ 2(1-\alpha) H_H
\label{H_bulk}
\end{align}
where $H_F$ is the bulk Hamiltonian for Fredkin model Eq.\eqref{H_Fredkin},  
and $H_H =-\sum_i \vec\sigma_i\cdot \vec\sigma_{i+1}$.  
We use the same boundary terms as before, which favors $|\uparrow\rangle$ on the left boundary and $|\downarrow\rangle$ on the right boundary. 
When $\alpha=1$, we recover the Fredkin model Eq.\eqref{H_Fredkin}, while for $\alpha=0$, the Hamiltonian
reduces to the isotropic ferromagnetic Heisenberg interaction. Therefore, by varying the coefficient $\alpha$ we 
can study the crossover from the Fredkin to the Heisenberg model. Away from $\alpha=1$, 
the Hamiltonian cannot be simply expressed in terms of projectors (except if $\alpha=0$), and thus the analytical form of the groundstate is currently unknown. We will use DMRG to compute various properties for the groundstate, and then study the low lying excited states. 

\begin{figure}[hbt]
\centering
 \subfigure[]{\label{fig:Fredkin_sz} \includegraphics[width=.46\textwidth]{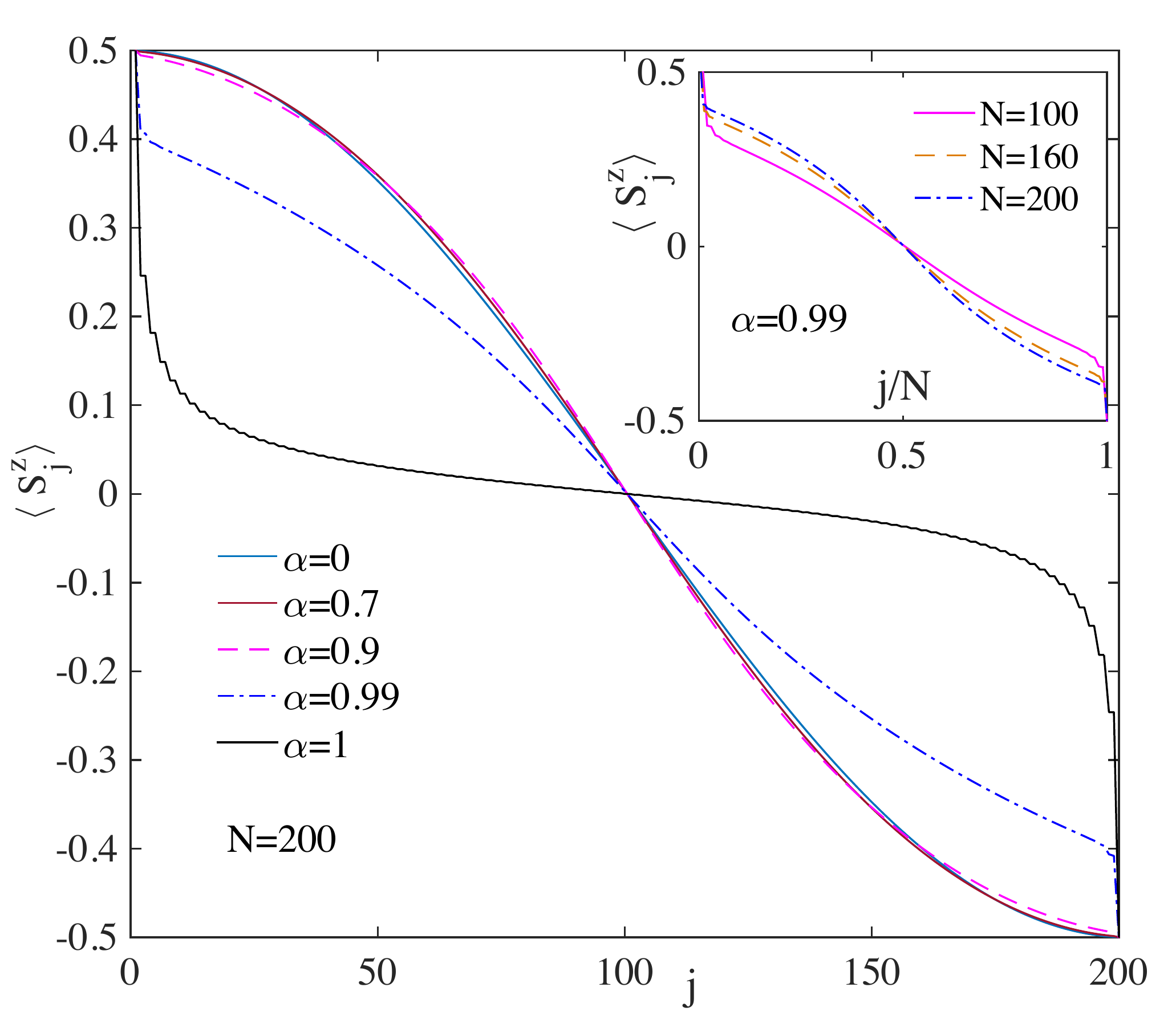}}   
 \subfigure[]{\label{fig:Fredkin_EE} \includegraphics[width=.46\textwidth]{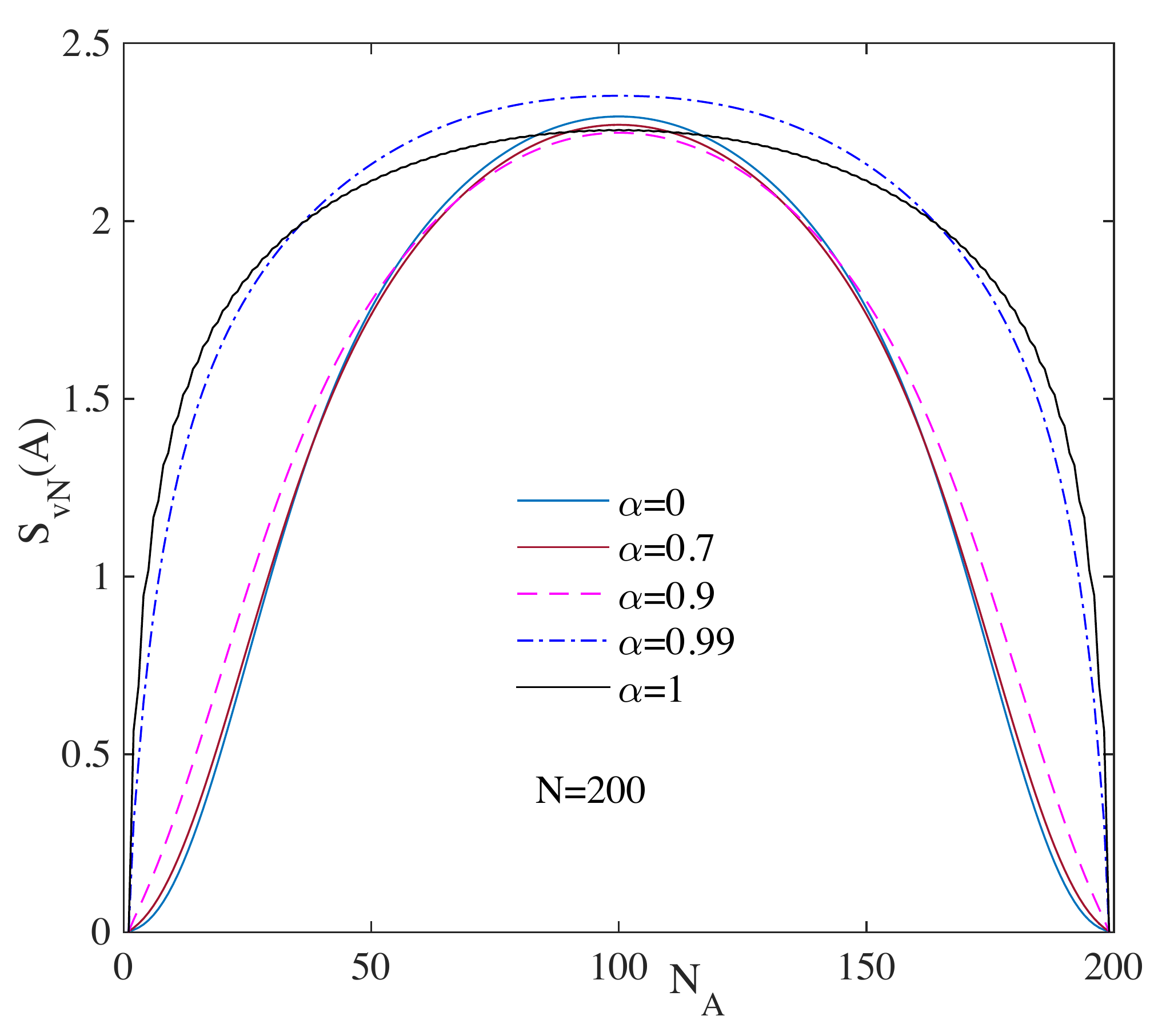}}
\caption{
(a) $\langle S^z_j\rangle$ vs site number $j$ for various $\alpha$ in the groundstate of the Fredkin-Heisenberg model obtained using DMRG. In the inset, we compare $\langle S^z_j\rangle$ for various system size at $\alpha=0.99$.
(b) $S_{\rm vN}(A)$ for various $\alpha$ in the Fredkin-Heisenberg model, where subsystem $A$ corresponds to the
first $N_A$ sites of the chain. {\color{black}The dotted line is the analytical result which matches the numerical result for $\alpha=1$.}
} 
\label{fig:Fre_EE_sz}
\end{figure}

For the Hamiltonian defined in Eq.\eqref{H_bulk}, at $\alpha=0$, if there is no extra boundary term, we can easily write down  
one groundstate: $|\Psi_0\rangle = |\uparrow\uparrow\ldots\uparrow\rangle$, which is a product state. 
Since the $\alpha=0$ model is isotropic, there are also degenerate ferromagnetic states in other directions. 
The low energy excited state can then be described by spin wave excitations with a dynamical exponent $z=2$. 
However, once we turn on the boundary term, the situation becomes more complicated. 
The boundary term favors $|\!\uparrow\rangle$ on the left boundary and $|\!\downarrow\rangle$ on the right boundary.
Instead of the groundstate being a naive product state with a sharp domain wall in it, 
we find that $\langle S^z_j\rangle$ changes continuously from $+1/2$ to $-1/2$ in order to lower the total energy, 
as shown in Fig.~\ref{fig:Fredkin_sz}.
The whole system forms a smooth 
domain wall which is a singlet, $S^z_{\rm tot}=0$. 
As shown in Fig.~\ref{fig:Fredkin_EE}, this state has a large EE, which increases as we increase the subsystem size $N_A$ (until $N_A=N/2$), 
and exceeds the area law (i.e.\ a constant in 1d) obeyed by gapped systems. This contributes to making the 
DMRG calculations time-consuming.    

We further study the groundstate at finite $\alpha$. In Fig.~\ref{fig:Fre_EE_sz}, we show the EE 
and $\langle S^z_j\rangle$ for the groundstate of the Fredkin-Heisenberg model at various $\alpha$ with $N=200$. We notice that as we increase $\alpha$ from 0 to $0.7$, both the EE (Fig.~\ref{fig:Fredkin_EE})  and $\langle S^z_j\rangle$ (Fig.~\ref{fig:Fredkin_sz}) barely change, suggesting that the groundstate is very close to the $\alpha=0$ one. These quantities show clearer deviations when $\alpha$ approaches 0.9. 
At $\alpha=0.99$ and for $N=200$, the EE and $\langle S^z_j\rangle$ differ from the small $\alpha$ and $\alpha=1$ cases.  
We examine the finite size effects in $\langle S^z_j\rangle$ at $\alpha=0.99$, the result is shown in the inset of Fig.~\ref{fig:Fredkin_sz}. As we increase $N$, $\langle S^z_j\rangle$ approaches the $\alpha=0$ result. Therefore we expect that in the thermodynamic limit, the groundstate at $\alpha=0.99$ is quite different from that at $\alpha=1$, suggesting that the Heisenberg interaction is a relevant perturbation to the Fredkin model.  

\begin{figure}[hbt]
\centering
 \subfigure[]{\label{fig:gap} \includegraphics[width=.46\textwidth]{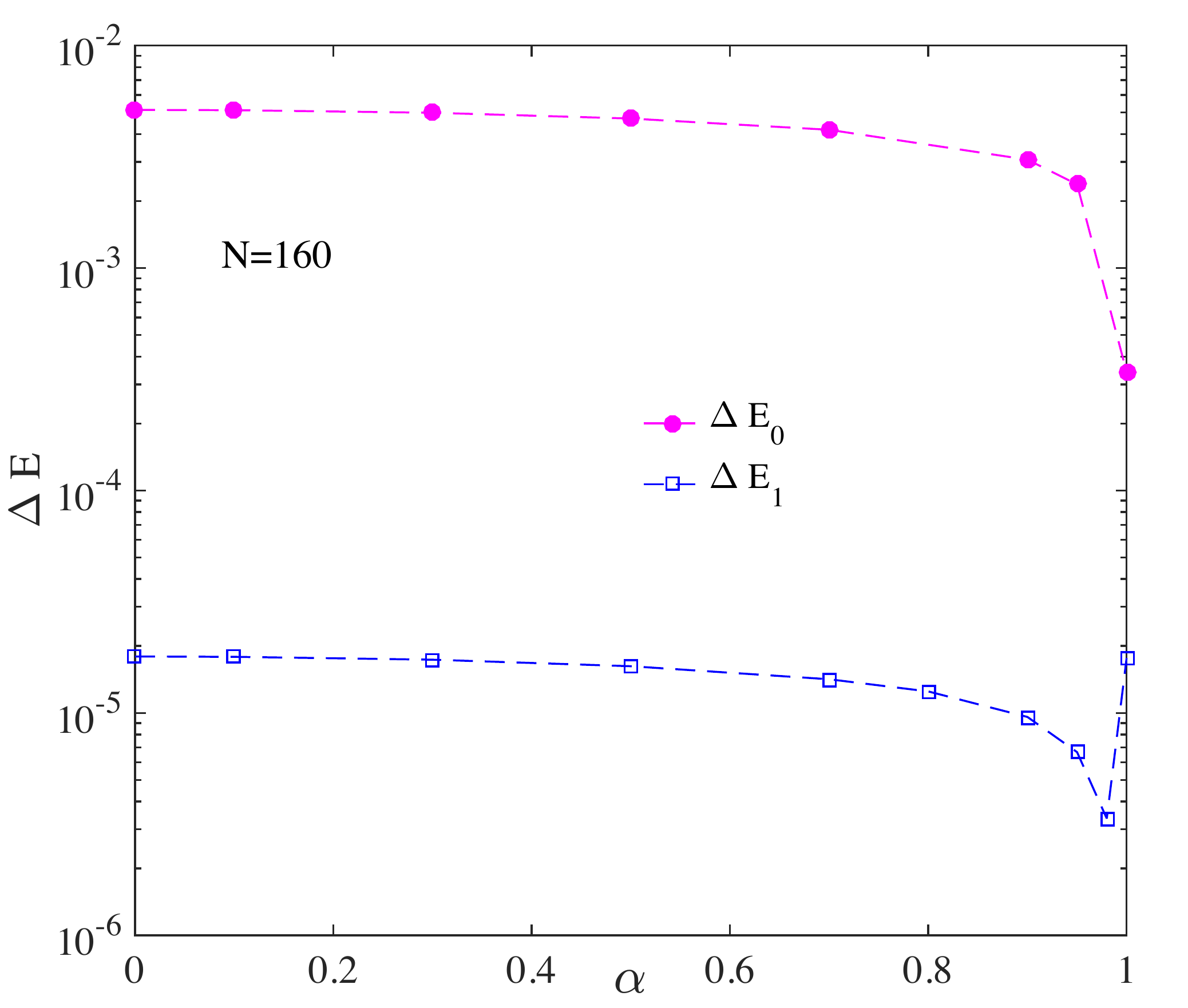}}   
 \subfigure[]{\label{fig:Fredkin_z} \includegraphics[width=.46\textwidth]{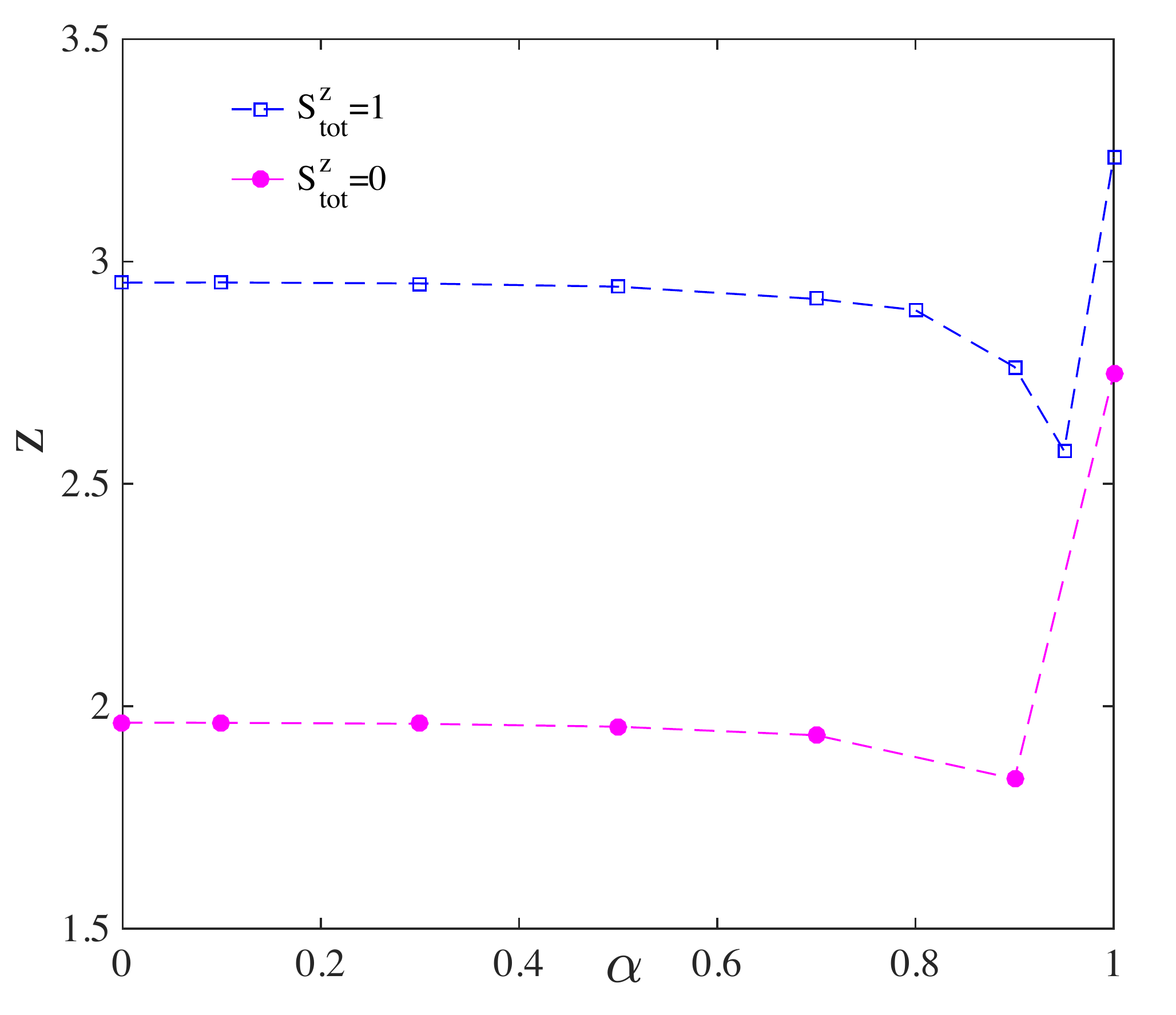}}
\caption{
(a) Energy gap vs $\alpha$ in the Fredkin-Heisenberg model obtained using DMRG with fixed $N=160$; the scale is log-linear. The blue squares are the energy gap $\Delta E_1$ between the true lowest excitation in $S^z_{\rm tot}=1$ sector and the groundstate. The pink circles give the energy gap $\Delta E_0$ between the lowest excitation in the $S^z_{\rm tot}=0$ sector and the groundstate.
(b) Dynamical exponent vs $\alpha$ in the Fredkin-Heisenberg model. The blue squares give $z$ for 
the lowest energy excitation, which has $S^z_{\rm tot}=1$.
The pink circles give $z_0$ for the lowest excitation in the $S^z_{\rm tot}=0$ sector.  
} 
\label{fig:Fre_excited}
\end{figure}

We now investigate the energy gap for Fredkin-Heisenberg model.  We notice that as long as $0\leq \alpha \leq 1$, 
the groundstate is always in $S^z_{\rm tot}=0$ sector with the first excited state in $S^z_{\rm tot}=\pm 1$ sector. 
Furthermore, the model remains gapless with the energy gap $\Delta E_1$ scales as $1/N^z$. 
As shown in Fig.~\ref{fig:gap}, for $N=160$, $\Delta E_1$ is decreasing as we increase $\alpha$ from zero. 
We observe an abrupt change around $\alpha=1$, at which point $\Delta E_1(\alpha=1)$ jumps to a larger value. 
In contrast, we do not observe the same abrupt change in the energy gap $\Delta E_0$ between the lowest excitation in
the $S^z_{\rm tot}=0$ sector and the groundstate. $\Delta E_0$ decreases continuously as we increase $\alpha$ from zero.   

Once we know the size dependence of the energy gap, we can extract the dynamical exponent $z$ by fitting to $1/N^z$.
We present the result for $z$ as a function of $\alpha$ in Fig.~\ref{fig:Fredkin_z} with  
the details for the DMRG calculations explained in Appendix~\ref{ap:dmrg}. 
For the lowest excitation,  
if $\alpha\leq 0.8$, the dynamical exponent is $z\simeq 3$, with small finite size corrections.    
This is the same as that for the Kawasaki (i.e.\ spin conserving) dynamics of the 1d Ising chain at low temperatures
compared with the exchanging coupling.\cite{Kawasaki1966, Cordery1981} The conservation law and the boundary effect slow down the dynamics and are responsible for the large $z$ here. As $\alpha$ approaches 1, we observe a dip in $z$, which eventually climbs back to its
$\alpha=1$ value, $z=3.23$.  
In order to assess the validity of this non-monotonic behavior in the thermodynamic limit,
we have analyzed finite-size dynamical exponent $z(N)$ (introduced above). The analysis, 
presented in Fig.~\ref{fig:Fre_z_finite} of Appendix~\ref{ap:dmrg},
suggests that non-negligible finite-size corrections exist at $\alpha=0.9$ and $\alpha=0.95$, 
with $z(\infty)$ being is closer to 3 than in Fig.~\ref{fig:Fre_excited} (but still smaller than 3).   
The variation of $z$ as a function of a parameter in the system was also observed in the generalized Motzkin model \cite{short-prep}, 
and the classical (Kawasaki) non-equilibrium dynamics of an Ising spin chain, where $z$ was found to change from 2 to 3.2.\cite{Grynberg}  

We also explore the dynamical exponent $z_0$ for the first excited state in the $S^z_{\rm tot}=0$ sector, 
and find that $z_0$ is much smaller than $z$ (pink circles in Fig.~\ref{fig:Fredkin_z}). When $\alpha\leq 0.7$, $z_0\simeq 2$, 
which is indicative of diffusive dynamics in that sector. When $\alpha \approx 0.9$, $z_0$ dips to a value less than 2. Similar to $z$, 
we observe a discontinuity for $z_0$ around $\alpha=1$.

\subsubsection{Comparison with conformal-wavefunction quantum critical points in 2+1D}
We compare our results for the dynamics with what was found for the 2d lattice models studied in Ref.~\onlinecite{Isakov2011}. 
In particular, they considered the so-called quantum eight-vertex and six-vertex models. Both models have a parameter that can be varied such
that the groundstate remains conformally invariant but with equal time two-point functions that change continuously.  
For the quantum eight-vertex model without $U(1)$ symmetry, the quantum dynamics maps to a classical non-equilibrium universality class
in \emph{model A}, according to the classification by Hohenberg and Halperin.\cite{Hohenberg1977} As the the parameter is varied, 
the dynamical exponent changes non-monotonically and obeys $z > 2$.  On the other hand, the quantum six-vertex Hamiltonian has $U(1)$ symmetry,
and thus maps to the \emph{model B} universality class. This Hamiltonian can be described by the quantum Lifshitz model and always has $z=2$.
This is different from our 1d spin chain where the conservation law slows down the dynamics and leads to $z>2$. 

\subsubsection{Role of boundary terms} 
The boundary term is important in the Fredkin-Heisenberg model as it can change both the groundstate and dynamical exponent $z$. 
This is because there are a large number of low-lying states with similar energies and the boundary condition will 
project out some of these states.
To study such effects, we introduce 2 tuning parameters, $\beta$ and $\gamma$, in the boundary term:
\begin{align}
  H_{\rm bdy}=\frac{\beta}{2}(1-\sigma_1^z)+\frac{\gamma}{2}(1+\sigma_N^z)
  \label{H_bdy}
\end{align}
Similar boundary terms have been studied for the Fredkin model in Ref.~\onlinecite{Salberger2016}, with an emphasis on groundstate properties.    
When $\beta=\gamma=1$ we recover the conditions used in the rest of the present work; in that case the groundstate is highly entangled.  
In contrast, if we choose $\beta=-\gamma=1$, for $0\leq \alpha\leq 1$, the groundstate has all the spins pointing up which is a trivial product state. The lowest excited state is the spin wave excitation with one spin pointing down. 
In the language of non-equilibrium classical dynamics,
such a down spin will move diffusively on the lattice in the background of up spins, which leads to a dynamical exponent $z=2$. Actually, the explicit values of $\beta$ and $\gamma$ are not very important, as long as they are finite.  We summarize the effects of the boundary 
terms in Table \ref{table_sum}.  

\begin{table} 
\centering
\begin{tabular}{c|c|c|c|c}
 & groundstate & $\langle S^z_j\rangle$ in the bulk & $z$ & DMRG \\\hline
$\alpha=1, \quad\beta=\gamma>0$ & Dyck path, $S^z_{\rm tot}=0$ & $0$ & 3.23 & slow \\
$0\leq \alpha <1,\quad \beta=\gamma>0$ & highly entangled state, $S^z_{\rm tot}=0$ & $\langle S^z_j\rangle \in (-\frac{1}{2}, \frac{1}{2})$ & $z>2$ & slow \\
$0\leq \alpha \leq 1, \quad \beta=-\gamma>0$ & product state, $S^z_{\rm tot}=N/2$ & $\frac{1}{2}$ & $2$ & fast
\end{tabular}
\caption{The property of Hamiltonian in Eq.\eqref{H_bulk} with various boundary conditions. The last column refers to the convergence
speed of the DMRG algorithm. 
}\label{table_sum}
\end{table}



\section{Conclusion and Outlook}
\label{conclusion}
We have studied two quantum spin chains, the so-called Fredkin ($S=1/2$) and Motzkin ($S=1$) models, using a variety of methods: 
exact relations, field theory, and DMRG. We have found that the entangled groundstate of both Hamiltonians takes
the same form in the continuum, with a dimensionless parameter $\kappa$ that depends on the spin. The wavefunction
is expressed in terms of a continuum height field $\phi$ that acts like a ``gauge field'' for the spin: $S^z(x)=\partial_x\phi$.
The resulting wavefunction can be viewed as  the path integral representation of a quantum particle restricted 
to move on the positive half line, $\phi\geq 0$. Relying on this feature, we computed various properties for the groundstate and compared 
them with the lattice results for the Fredkin and Motzkin models. We showed that the groundstate has a large entanglement entropy 
but with zero mutual information between two disjoint intervals deep inside the system, suggesting that it is less entangled than
typical $(1+1)$D CFTs. The mutual information result is consistent with two-point correlation function of the local operator $S^z$, $\langle S^z(x_1)S^z(x_2)\rangle=0$. On the other hand, $\langle S^+_iS^-_j\rangle$ in both spin models saturates to a finite 
constant deep inside the bulk; this is consistent because $S^+_i S^-_j$ corresponds to a non-local string operator in the height representation.  


Based on these results, we found that deep inside the bulk
the groundstate wavefunction enjoys 
an emergent $(1+0)$ dimensional conformal-type symmetry, in the sense of conformal quantum mechanics. 
The Fredkin and Motzkin models can thus be considered as the lower dimensional analogues of
conformal quantum critical points, whose wavefunctions have a two dimensional spatial conformal symmetry.\cite{Ardonne-2004,Isakov2011,Fradkin-book}  

The approach discussed in this paper connects 1d RK states with a simple quantum mechanics problem. 
It can thus be used to construct other highly entangled RK states by introducing a potential term for the 
quantum mechanical particle, such as the exponential potential of Liouville quantum mechanics. 
It would be of interest to study these states and find lattice models that realize them.

A key motivation of this paper was to investigate the dynamical exponent for the Fredkin model. 
Following our previous DMRG results for the Motzkin model,\cite{short-prep} we performed large-scale DMRG 
for the Fredkin Hamiltonian. We found that the dynamical exponent for the lowest excitation with $S^z_{\rm tot}=\pm 1$ 
has $z=3.23$. 
We mapped the quantum dynamics to the classical non-equilibrium relaxation of the corresponding classical spin chain.
This gives a heuristic explanation of the large dynamical exponent in terms of the subdiffusive relaxation in the classical system. 
Moreover, we found the higher energy $S^z_{\rm tot}=0$ excitation has a dynamical exponent $z_0=2.76$, 
an indication of  multiple dynamics in the Fredkin model, just as what we previously found for the Motzkin model.\cite{short-prep} 
In fact, both the $z$ and $z_0$ dynamical exponents are very close to those found in the Motzkin model. 
It is tantalizing to speculate that both models can be described by the same effective field theory. We leave this interesting question for
future work.  

Finally, we explored the crossover from the Fredkin model to the ferromagnetic Heisenberg model as a function of a tuning parameter 
$\alpha$ in the Hamiltonian. Under the fixed boundary condition which favors up (down) spin on the left (right) end, 
we found using DMRG that the bulk Heisenberg interaction is a relevant perturbation and can drastically change the entire spectrum. 
Further study of the excited states indicates that the model remains gapless as we vary $\alpha$, and again shows multiple dynamics in different 
spin sectors. It would be interesting to have a better understanding for these phenomena in terms of a low energy theory.

We close by noting that finding a continuously varying dynamic critical exponent $z$ 
is by itself puzzling. In equilibrium classical systems, and in quantum theories with relativistic dynamics, continuously varying exponents occur when the system has an exactly marginal operator \cite{Kadanoff1977}, which on itself is a highly uncommon situation. Classical critical dynamics with non-trivial values of the dynamical exponent at the non-trivial fixed points of the classical equilibrium systems are also quite common \cite{Hohenberg1977}. On the other hand, except for the models studied here  (and in Refs.\cite{Grynberg,Isakov2011,short-prep}), 
there are very few other known cases of theories with varying values of $z$. 
It would be interesting to understand the mechanism(s) for continuously varying dynamic critical exponents. 

\begin{acknowledgements}  
We thank L.~Balents, J.~Cardy, L.~Dell'Anna, A.~Ludwig, R.~Movassagh, S.~Sachdev, M.~Stoudenmire  and X.~Yu for useful discussions. XC was supported by a postdoctoral fellowship from the the Gordon and Betty Moore Foundation,
under the EPiQS initiative, Grant GBMF4304, at the Kavli Institute for Theoretical
Physics.
This work was supported in part by the US  National Science Foundation through grant DMR 1408713 at the University of Illinois (EF).  
WWK was funded by a Discovery Grant from NSERC, and by a Canada Research Chair. 
The DMRG simulations were performed using the ITensor package (v2). We acknowledge support from the Center for Scientific Computing from the CNSI, MRL: an NSF MRSEC (DMR-1121053).
\end{acknowledgements}

\appendix
\section{Some combinatorics} 
\label{Dyck_Motzkin}
\subsection{Dyck paths}
\begin{figure}
\centering
\includegraphics[width=.52\textwidth]{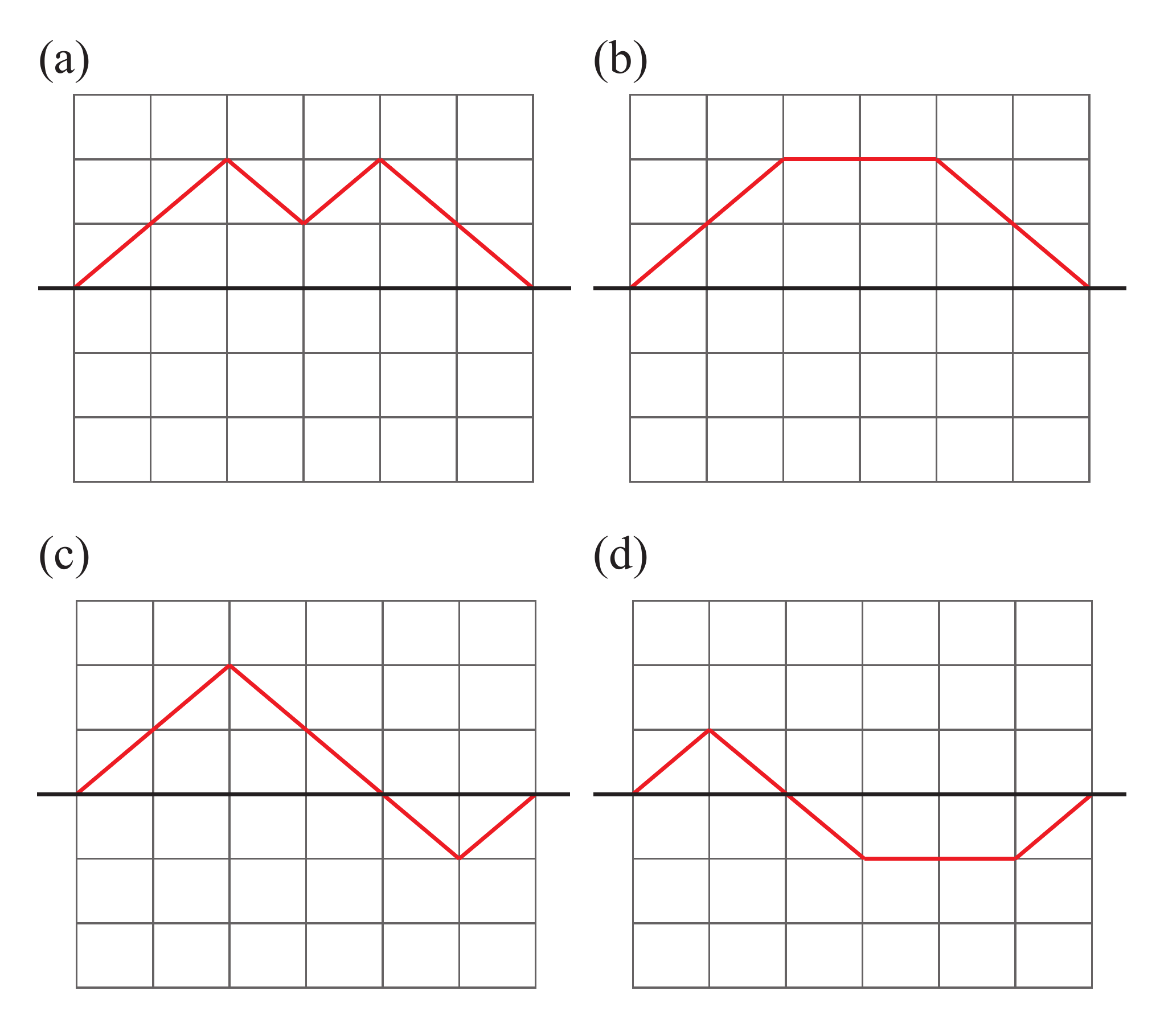}
\caption{(a) is an example for Dyck path, both (a) and (b) belong to Motzkin path. (a),(c) represent all the unrestricted paths in spin $1/2$ system, all the plots are the allowed paths in spin $1$ system.}
\label{fig:dyck}
\end{figure}

A Dyck path is a path which starts at $(x,y)=(0,0)$ and ends at $(x,y)=(N,0)$, and consists of two types of moves: diagonal up $(1,1)$ and diagonal down $(1,-1)$.  The Dyck path needs to satisfy $y\geq 0$ and when $N$ is an even number, the number of allowed Dyck paths is given by the Catalan number
\begin{align}
D_{N,0,0}=\frac{1}{\frac{N}{2}+1}\begin{pmatrix} N \\ \frac{N}{2}\end{pmatrix}
\end{align}
One simple example of Dyck path is shown in Fig.~\ref{fig:dyck} (a). 

If we consider a more general case and let the walker start at $(x,y)=(0,m_1)$ and end at $(x,y)=(N,m_2)$ with $m_1\geq 0$ and $m_2\geq 0$. The number of the Dyck path above the  upper half-plane is
\begin{align}
D_{N,m_1,m_2}=\begin{pmatrix} N \\ \frac{N+|m_2-m_1|}{2}  \end{pmatrix} - \begin{pmatrix} N \\ \frac{N+(m_2+m_1)}{2}+1  \end{pmatrix}
\end{align}
When $m_1=0$ and $m_2=m$, the above expression can be simplified to
\begin{align}
D_{N,0,m}=\frac{m+1}{N+1}\begin{pmatrix} N+1 \\ \frac{N-m}{2} \end{pmatrix}
\end{align}
In the limit $N\to\infty$, we have 
\begin{align}
\begin{pmatrix} N \\ \frac{N+m}{2}  \end{pmatrix}\approx \frac{2^N}{\sqrt{\pi N/2}}e^{-\frac{m^2}{2N}}
\end{align}
where we assume $m\ll N$ and we also use Stirling's approximation
\begin{align}
n!\approx \sqrt{2\pi n}\left(\frac{n}{e}\right)^n.
\end{align}
At $x=N_A$ between $0$ and $N$, the probability of a Dyck path with height $y=m$ is given by
\begin{align}
P(N_A,m)=\frac{D_{N_A,0,m}D_{N-N_A,m,0}}{\sum_m D_{N_A,0,m}D_{N-N_A,m,0}}=\sqrt{\frac{2}{\pi}} \left(\frac{N}{N_A(N-N_A)}\right)^{3/2} m^2e^{-\frac{N}{N_A(N-N_A)}\frac{m^2}{2}}
\end{align}  

\subsection{Motzkin paths}
For a Motzkin path, the walker still starts at $(x,y)=(0,0)$ and ends at $(x,y)=(N,0)$ with the path always satisfying $y\geq 0$. Apart from the diagonal up $(1,1)$ and diagonal down $(1,-1)$ in the Dyck path, horizontal $(1,0)$ move is also allowed. The total number of Motzkin path is called Motzkin number and is equal to 
\begin{align}
M_{N,0,0}=\sum_{k=0}^{L}\begin{pmatrix} N \\ k \end{pmatrix} D_{N-k,0,0}
\end{align}
where the binomial coefficient in front of Catalan number $D_{N-k,0,0}$ is the number of allowed horizontal steps. Two simple examples of Motzkin path are shown in Fig.~\ref{fig:dyck} (a) (b).

We can also consider a more general case with  the walker starts at $(x,y)=(0,m_1)$ and ends at $(x,y)=(N,m_2)$. The number of allowed Motzkin paths is
\begin{align}
M_{N,m_1,m_2}=\sum_{k=0}^{N-|m_2-m_1|}\begin{pmatrix} N \\ k \end{pmatrix} D_{N-k, m_1,m_2}
\end{align}
When $m_1=0, m_2=m$, the above expression can be simplified to
\begin{align}
M_{N,0,m}=\frac{m+1}{N+1}\sum_{i\geq 0}\begin{pmatrix} N+1 \\ N-2i-m, i, i+m+1  \end{pmatrix}
\label{M_0m}
\end{align}
where $2i=N-k-|m_2-m_1|$ and $\begin{pmatrix} N \\ x,  y, z \end{pmatrix}\equiv \frac{N!}{x!y!z!}$ is the trinomial coefficient which takes the maximum value at $x=y=z=N/3$.\cite{Movassagh2017} In the limit $N\to\infty$,  we expand the trinomial coefficient around this saddle point solution by using Stirling's approximation, 
\begin{align}
\begin{pmatrix} N \\ \frac{N+a}{3}, \frac{N+b}{3}, \frac{N+c}{3} \end{pmatrix}\approx \frac{3^{N+\frac{3}{2}}}{2\pi N}e^{-\frac{3}{2N}(a^2+b^2+c^2)}
\end{align}
with $a+b+c=0$. For the trinomial coefficient in Eq.\eqref{M_0m}, notice that we can define $N-2i-m=\frac{N}{3}+a$, $i=\frac{N}{3}+b$ and $i+m=\frac{N}{3}+c$, therefore the sum in this expression can be replaced by the integration around this saddle point $i=N/3$ and we have 
\begin{align}
M_{N,0,m}\approx \frac{3^{N+1/2}m}{2\sqrt{\pi} N^2}e^{-\frac{3m^2}{4N}}.
\end{align}

At $x=N_A$ between $0$ and $N$, the probability of a Motzkin path with height $y=m$ is given by
\begin{align}
P(N_A,m)=\frac{M_{N_A,0,m}M_{N-N_A,m,0}}{\sum_m M_{N_A,0,m}M_{N-N_A,m,0}}=\sqrt{\frac{2}{\pi}} \left(\frac{3N}{2N_A(N-N_A)}\right)^{3/2} m^2e^{-\frac{N}{N_A(N-N_A)}\frac{3m^2}{4}}
\end{align}

\section{DMRG calculations}
\label{ap:dmrg}

We have numerically calculated the energy gap between the groundstate and the lowest energy excited states $\Delta E$ 
by using both exact diagonalization (ED) and density matrix renormalization group (DMRG). 
We use ED for small systems $N\leq 10$ as a benchmark, 
and we perform large-scale DMRG calculations using the open-source C++ library ITensor. 
For the Motzkin model, and its generalization, the DMRG results have been presented in Ref.~\onlinecite{short-prep}. 

For the Motzkin and Fredkin models, $\Delta E$ scales as $1/N^z$ for sufficiently large $N$. 
Since the dynamical exponent $z$ is large and close to 3, 
it makes the DMRG more difficult, especially
if high precision is required. As a first step, we compare the groundstate energy and the von Neumann EE obtained via DMRG 
with the analytical results and find that they agree precisely.
Then, in order to determine the lowest excited state, which we find has $S^z_{\rm tot}=\pm 1$, a large number of sweeps 
are used to ensure that the energy is properly converged.  
The energy deviation between the last two successive sweeps is around $10^{-12}$. 

\begin{figure}[hbt]
\centering
 \subfigure[]{\label{fig:Mot_corr} \includegraphics[width=.47\textwidth]{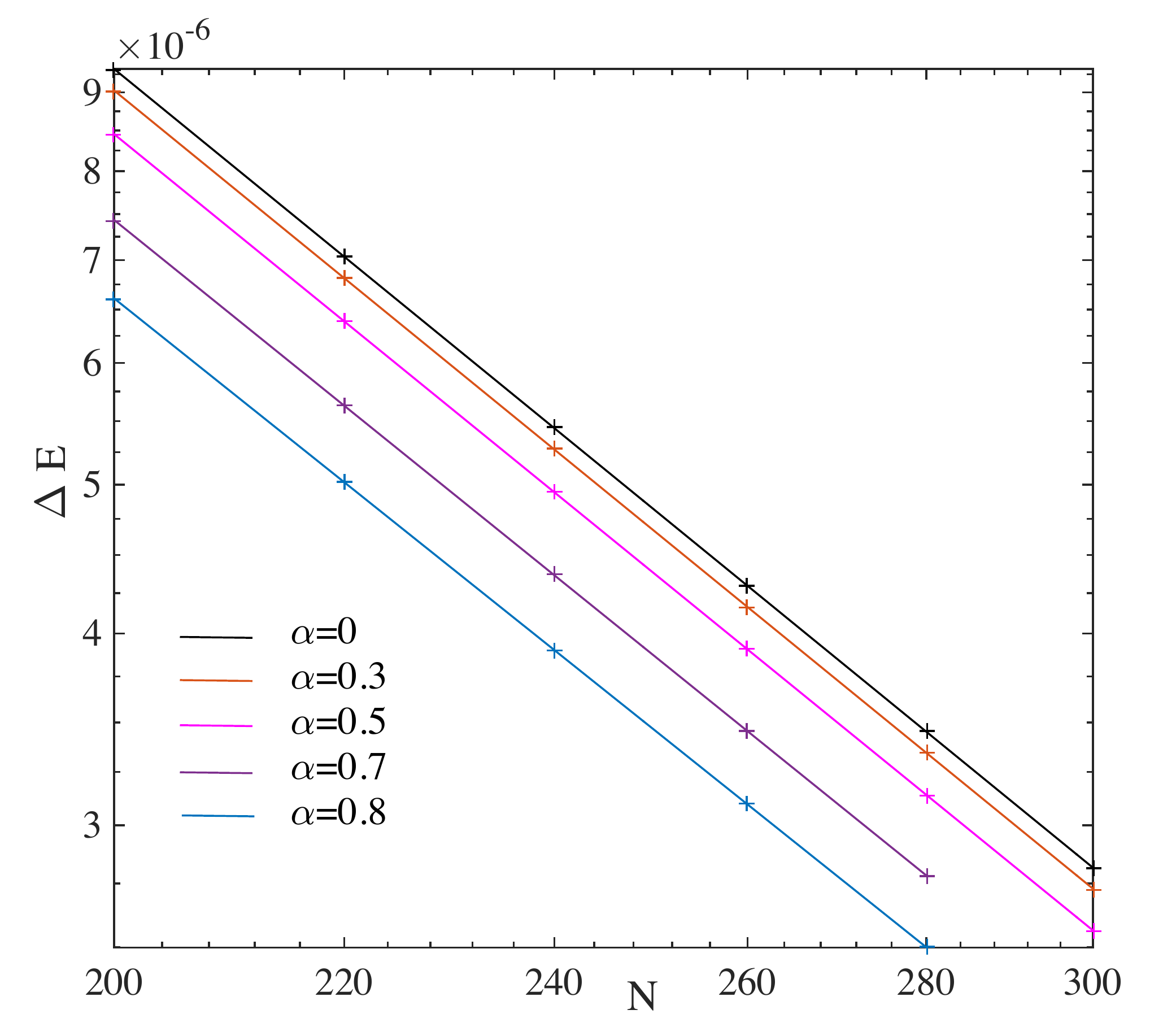}}
 \subfigure[]{\label{fig:Fre_corr} \includegraphics[width=.45\textwidth]{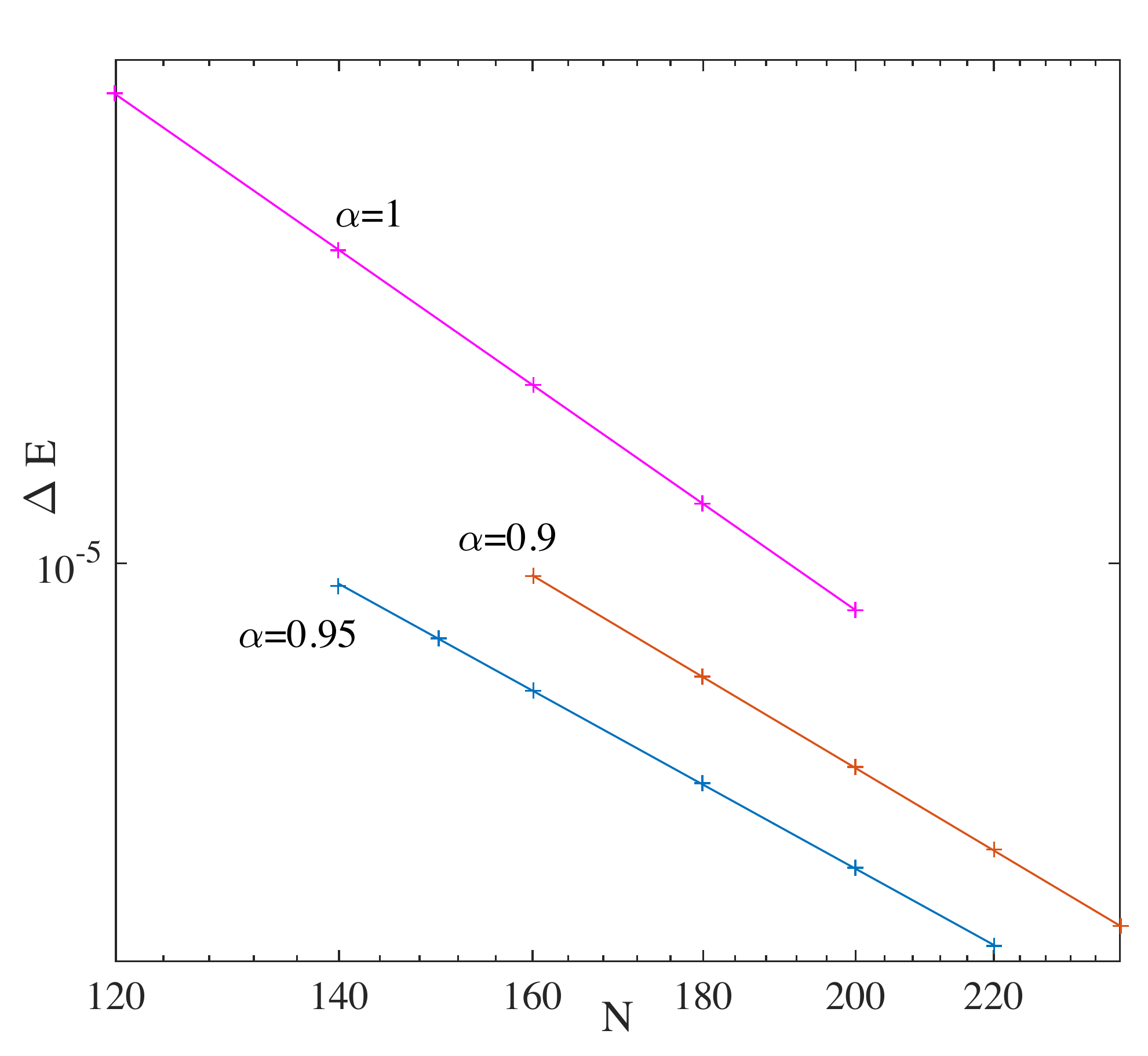}}   
\caption{
(a) Log-log plot of the energy gap $\Delta E$ versus system size for $0\leq \alpha\leq 0.8$. The lines are fits to $\Delta E\propto N^{-z}$, with the range of $N$ given in the caption of Table~\ref{table_z}. (b)  Log-log plot of the energy gap $\Delta E$ versus system size for $0.9\leq \alpha\leq 1$. The lines are fits to $\Delta E\propto N^{-z}$, with the range of $N$ given in the caption of Table~\ref{table_z}.} 
\label{fig:E_vs_N}
\end{figure}

We calculate the energy for the lowest excited state in $S^z_{\rm tot}=1$ sector. We show the energy gap $\Delta E$ between 
the groundstate and the lowest excited state in Fig.~\ref{fig:E_vs_N} as a function of system size. 
We fit $\Delta E\propto 1/N^z$ using the data points in the range $120\leq N\leq 200$ in order to minimize finite-size effects that appear at smaller $N$. 
Moreover, we calculate the lowest excited state with $S^z_{\rm tot}=0$ and show the dynamical exponent $z_0$ in Table \ref{table_z}.  
We consider smaller system sizes $100\leq N\leq 160$ because 
the calculation of the excited state in the same spin sector as the groundstate  is time-consuming. 

For the Fredkin-Heisenberg model with $\alpha<1$, we do not know the analytical expression for the groundstate.  
To compute the energy gap, we need to calculate the energies of both the groundstate and excited state numerically. 
We then use the same method as above to extract both $z$ and $z_0$.  The fit for $\Delta E\propto 1/N^z$ is shown in Fig.~\ref{fig:E_vs_N} and the detail of the 
range for $N$ is listed in Table~\ref{table_z}. 

\begin{table}[htbp]
\centering
\begin{tabular}{c||c|c|c|c}
$\alpha$ & $z$   & $z(\infty)$  & $z_0$   & $z_0(\infty)$ \\ \hline \hline
0     & $2.95$  & $3.00$ & $1.96$   & 2.00\\ 
0.1  & $2.95$  & $3.00$ & $1.96$   & 2.00\\
0.3  & $2.95$  & $3.00$ & $1.96$   & 2.00\\
0.5  & $2.94$  & $3.00$ & $1.95$   & 2.00\\
0.7  & $2.92$  & $3.00$ & $1.93$   &1.99\\
0.8  & $2.89$  & $2.99$ &               &\\
0.9  & $2.74$  & $2.98$ & $1.82$   & \\
0.95& $2.50$  & $2.74$ &               &\\
1     & $3.23$  & $3.19$ & $2.76$   & 2.69\\
\end{tabular}
\caption{The dynamical exponents of the Fredkin-Heisenberg spin chain as a function of $\alpha$, both in the $S^z_{\rm tot}=1$ and $S^z_{\rm tot}=0$ sectors. The lowest energy excited state has $S^z_{\rm tot}\!=\! 1$. For $z$, when $0\leq \alpha\leq 0.5$, the range of $N$ is $200-300$, when $\alpha=0.7, 0.8$, the range of $N$ is $200-280$, when $\alpha=0.9$, the range of $N$ is $160-240$, when $\alpha=0.95$, the range of $N$ is $140-220$, when $\alpha=1$, the range of $N$ is $120-200$. For $z_0$, when $0\leq \alpha\leq 0.9$, the range of $N$ is $140-200$, when $\alpha=1$, the range of $N$ is $100-160$. 
We have not computed $z_0(\infty)$ for $\alpha=0.9$ since $z_0(N)$ does not scale linearly with $1/N$.}
\label{table_z} 
\end{table}

\begin{figure}
\centering
\includegraphics[width=.5\textwidth]{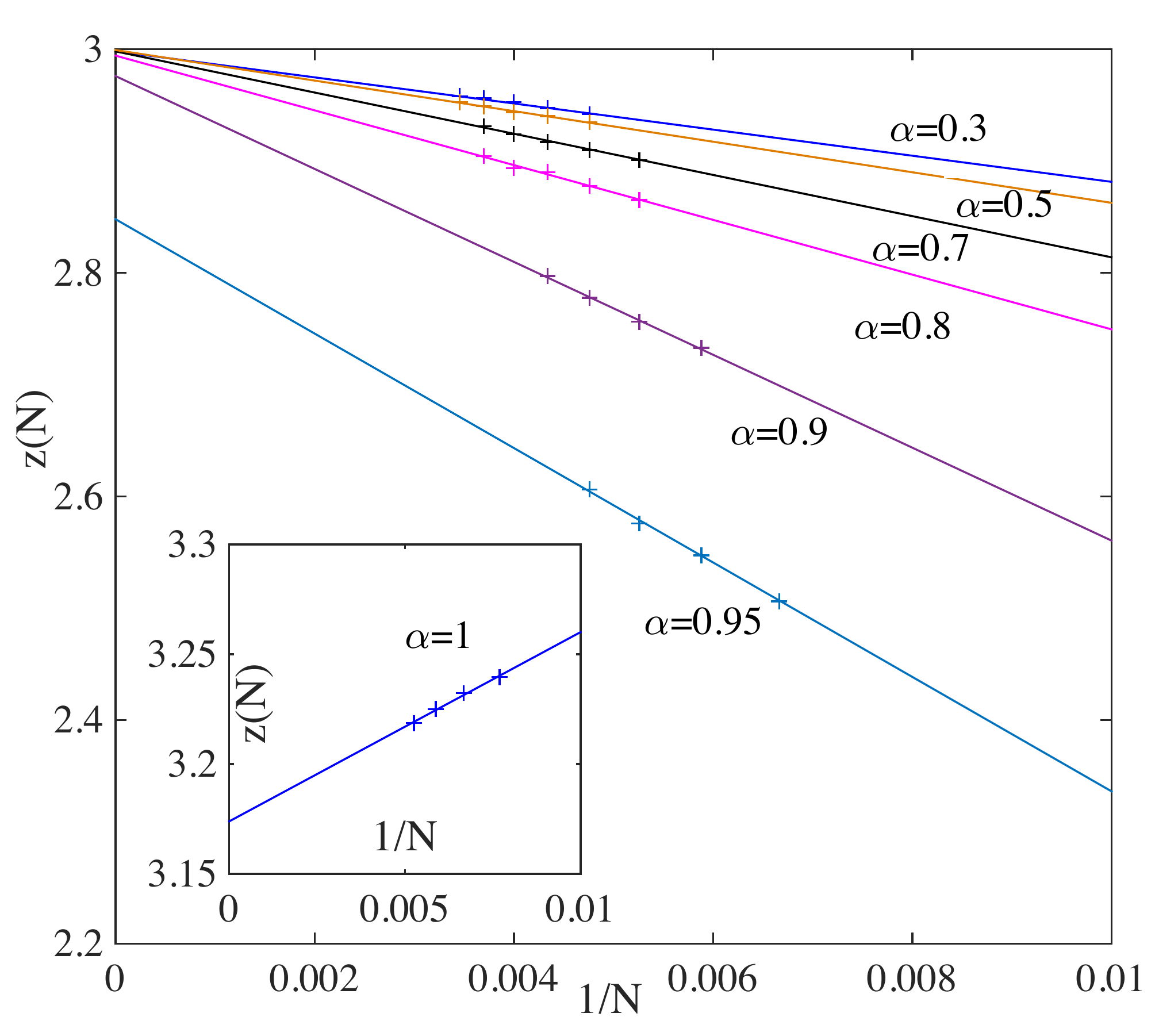}
\caption{Finite-size dynamical exponent $z(N)$ vs chain length $N$ for various $\alpha$ in the Fredkin-Heisenberg model. }
\label{fig:Fre_z_finite}
\end{figure}

In Fig.~\ref{fig:Fre_z_finite}, we show an alternate method to estimate $z$ in the thermodynamic limit. 
For 2 consecutive values of $N$, $N_1<N_2$, we evaluate the finite size exponent at the midpoint:
\begin{align}
  z \!\left(\frac{N_1+N_2}{2}\right) = - \frac{\ln(\Delta E_2/\Delta E_1)}{\ln (N_2/N_1)}  
\end{align}
where $\Delta E_i$ is the gap for system size $N_i$.
The dependence of $z(N)$ on $N$ is shown in Fig.~\ref{fig:Fre_z_finite} for different values of the coupling $c$. 
We notice that the variation of $z(N)$ with $N$ is small when $\alpha\leq 0.7$, indicating that finite size effects are small in this regime. 
As we increase $\alpha$, finite size effects for $z(N)$  
become larger and we notice that $z(N)$ roughly scales as $1/N$. In order to get an estimate for $z$ in the thermodynamics limit,
we use a $1/N$ fit to extract $z(N\to\infty)$. 
We compare the corresponding results with $z$ obtained from the fits described above 
in Table \ref{table_z}. 
It is important to emphasize that the $1/N$ extrapolation for $z(N)$ may not be accurate at $c>0.7$, and is only a crude estimate for the true $z$.
For instance, the slope could change at larger $N$, or the data could deviate from $1/N$ scaling.

\bibliographystyle{apsrev4-1}
\bibliography{Fredkin}{}     

\end{document}